\documentclass[12pt]{article}
\usepackage{amsmath,amssymb}
\usepackage{epsfig,graphics,graphicx}
\usepackage{color,multicol}
\usepackage{setspace}
\usepackage{natbib}
\usepackage{multirow}
\usepackage{color, colortbl}
\definecolor{Gray}{gray}{0.9}
\usepackage{graphicx,psfrag,epsf}
\usepackage{enumerate}

\newcommand{\blind}{0}

\begin{document}

\def\spacingset#1{\renewcommand{\baselinestretch}%
{#1}\small\normalsize} \spacingset{1}

%%%%%%%%%%%%%%%%%%%%%%%%%%%%%%%%%%%%%%%%%%%%%%%%%%%%%%%%%%%%%%%%%%%%%%%%%%

\if0\blind
{
  \title{\bf Bayesian Nonparametric Modeling for Multivariate Ordinal Regression}
  \author{Maria DeYoreo\thanks{M. DeYoreo (maria.deyoreo@stat.duke.edu) is 
Postdoctoral Researcher, Department of Statistical Science, Duke University. This research 
is part of the Ph.D. dissertation of M. DeYoreo, completed at University of California, Santa Cruz.}\hspace{.2cm}\\
    Department of Statistical Science, Duke University\\
    and \\
    Athanasios Kottas\thanks{A. Kottas 
(thanos@ams.ucsc.edu) is Professor of Statistics, Department of Applied Mathematics 
and Statistics, University of California, Santa Cruz. 
} \\
    Department of Applied Mathematics and Statistics, \\ University of California, Santa Cruz}
  \maketitle
} \fi

\if1\blind
{
  \bigskip
  \bigskip
  \bigskip
  \begin{center}
    {\LARGE\bf Title}
\end{center}
  \medskip
} \fi

\bigskip
\begin{abstract}
Univariate or multivariate ordinal responses are often assumed to arise from a latent 
continuous parametric distribution, with covariate effects which enter linearly. 
We introduce a Bayesian nonparametric modeling approach for univariate and multivariate 
ordinal regression, which is based on mixture modeling for the joint distribution of latent 
responses and covariates. The modeling framework enables highly flexible inference for 
ordinal regression relationships, avoiding assumptions of linearity or additivity in the 
covariate effects. In standard parametric ordinal regression models, computational challenges 
arise from identifiability constraints and estimation of parameters requiring nonstandard 
inferential techniques.  A key feature of the nonparametric model is that it achieves inferential 
flexibility, while avoiding these difficulties.  In particular, we establish full support of the 
nonparametric mixture model under fixed cut-off points that relate through discretization 
the latent continuous responses with the ordinal responses. The practical utility of the modeling 
approach is illustrated through application to two data sets from econometrics, an example 
involving regression relationships for ozone concentration, and a
multirater agreement problem. 
Supplementary materials related to computation are available online.

\end{abstract}

\noindent%
\vfill

\newpage
\spacingset{1.45} % DON'T change the spacing!
\section{Introduction}
\label{sec:intro}

Estimating regression relationships for univariate or multivariate ordinal responses is a key problem in many 
application areas. Correlated ordinal data arise frequently in the social sciences, for instance, survey
respondents often assign ratings on ordinal scales (such as ``agree", ``neutral", or 
``disagree") to a set of questions, and the responses given by a single rater are correlated. Ordinal data is 
also commonly encountered in econometrics, as rating agencies (such as Standard and Poor's) use an ordinal 
rating scale. A natural way to model data of this type is to envision each ordinal variable as representing 
a discretized version of an underlying latent continuous random variable. In particular, the commonly used 
(multivariate) ordinal probit model results when a (multivariate) normal distribution is assumed for the 
latent variable(s).

Under the probit model for a single ordinal response $Y$ with $C$ categories, and covariate vector 
$\boldsymbol{x}$, $\mathrm{Pr}(Y\leq m\mid \boldsymbol{x})=$
$\Phi(\gamma_m-\boldsymbol{x}^T\boldsymbol{\beta})$, for $m=1,\dots,C$. Here,  $-\infty=\gamma_0<\gamma_1<\dots<\gamma_{C-1}<\gamma_C=\infty$ are cut-off points for 
the response categories, where typically $\gamma_1=0$ for identifiability. \citet{albert} have shown 
that posterior simulation is greatly simplified by augmenting the model with latent variables.  
In particular, assume that the ordinal response $Y$ arises from a latent continuous response $Z$, 
such that $Y=m$ if and only if $Z\in (\gamma_{m-1},\gamma_m]$, for $m=1,\dots,C$, and $Z\mid \boldsymbol{\beta}\sim \mathrm{N}(\boldsymbol{x}^T\boldsymbol{\beta},1)$, which yields $\mathrm{Pr}(Y=m\mid\boldsymbol{x})=$
$\int_{\gamma_{m-1}}^{\gamma_m}\mathrm{N}(z \mid \boldsymbol{x}^T\boldsymbol{\beta},1) \text{d}z$.

The multivariate probit model for binary or ordinal responses generalizes the probit model to 
accommodate correlated ordinal responses $\boldsymbol{Y}=$ $(Y_{1},...,Y_{k})$, where 
$Y_{j} \in$ $\{ 1,...,C_{j} \}$ for $j=1,...,k$, using a multivariate normal distribution for the 
underlying latent variables $\boldsymbol{Z}=$ $(Z_{1},...,Z_{k})$.
To obtain an identifiable model, restrictions must be imposed on the covariance matrix 
$\boldsymbol{\Sigma}$ of the multivariate normal distribution for $\boldsymbol{Z}$. 
One way to handle this is to restrict the covariance matrix to be a correlation matrix, 
which complicates Bayesian inference since there does not exist a conjugate prior for 
correlation matrices. \citet{chibgreen2} discuss inference under this model, using a 
random walk Metropolis algorithm to sample the correlation matrix. \citet{liu2001} and 
\citet{imai} use parameter expansion with data augmentation \citep{liu} to expand 
the parameter space such that unrestricted covariance matrices may be sampled.
%and a one-to-one mapping is used to imply a set of draws for correlation matrices. 
\citet{talhouk} work with a sparse correlation matrix arising from conditional independence 
assumptions, and use a parameter expansion strategy to expand the correlation matrix 
into a covariance matrix, which is updated with a Gibbs sampling step. 
%
%The multivariate ordinal probit model can be applied when $\boldsymbol{Y}$ is an ordinal vector, and in terms of latent variables, %assumes $\boldsymbol{Z}\mid\boldsymbol{\beta}\sim %\mathrm{N}(\boldsymbol{x}^T\boldsymbol{\beta},\boldsymbol{\Sigma})$, %with each element of $\boldsymbol{Y}$ determined %by the corresponding element of $\boldsymbol{Z}$, so that $Y_j=m$ if and only if %$\gamma_{j,m-1}<Z_j\leq \gamma_{j,m}$, %for $j=1,\dots,k$. 
%To avoid the issue of constrained covariance matrices in the multivariate ordinal probit model, 
%
\citet{webb} reparameterize $\boldsymbol{\Sigma}$ in a way that enables fixing its 
diagonal elements without sacrificing closed-form full conditional distributions. 
\citet{lawrence} use a parameter expansion technique, in which the parameter space includes 
unrestricted covariance matrices, which are then normalized to correlation matrices.  
In addition to the challenges arising from working with correlation matrices, the setting with 
multivariate ordinal responses requires estimation for the cut-off points, which are typically 
highly correlated with the latent responses.

The assumption of normality on the latent variables is restrictive, especially for data which contains 
a large proportion of observations at high or low ordinal levels, and relatively few observations at 
moderate levels. As a consequence of the normal distribution shape, there are certain limitations
on the effect that each covariate can have on the marginal probability response curves. In particular, 
$\mathrm{Pr}(Y_{j}=1 \mid x)$ and $\mathrm{Pr}(Y_{j}=C_{j} \mid x)$ are monotonically 
increasing or decreasing as a function of covariate $x$, and they must have the opposite type of 
monotonicity. The direction of monotonicity changes exactly once in moving from category 
$1$ to $C_{j}$ (referred to as the single-crossing property). In addition, the relative effect of 
covariates $r$ and $s$, i.e., the ratio of $\partial \mathrm{Pr}(Y_{j}=m\mid \boldsymbol{x})/\partial x_r$ 
to $\partial \mathrm{Pr}(Y_{j}=m\mid \boldsymbol{x})/\partial x_s$, is equal to the ratio of the 
$r$-th and $s$-th regression coefficients for the $j$-th response, which does not depend on 
$m$ or $\boldsymbol{x}$. That is, the relative effect of one covariate to another is the same for every 
ordinal level and any covariate value. We refer to \citet{boes} for further discussion of such properties.

Work on Bayesian nonparametric modeling for ordinal regression is relatively limited, 
particularly in the multivariate setting. In the special case of binary regression, 
there is only one probability response curve to be modeled, and this problem has received 
significant attention. Existing semiparametric approaches involve relaxing the normality 
assumption for the latent response \citep[e.g.,][]{newton,basu}, while others have targeted 
the linearity assumption for the response function \citep[e.g.,][]{mukho, walker, choudhuri}. 
For a univariate ordinal response, \citet{chibgreen} assume that the latent response arises from 
scale mixtures of normals, and the covariate effects to be additive upon transformation by cubic 
splines. This allows nonlinearities to be obtained in the marginal regression curves, albeit under 
the restrictive assumption of additive covariate effects. 
%Moreover, there are aspects of the spline-based approach such as prior specification and 
%choice of location and number of knots that make implementing the model non-trivial. 
\citet{gill} extend the ordinal probit model by introducing subject-specific random effects terms
modeled with a Dirichlet process (DP) prior.

\cite{chen} model the latent variables for correlated ordinal responses with scale mixtures of normal 
distributions, with means linear on the covariates.  Related, \citet{BaoHanson} assume the latent variables arise from mixtures of linear multivariate probit regressions, mixing on the regression coefficients and covariance matrices in the multivariate normal distribution. The latter modeling approach is an extension of the work in \citet{kottas}, where, in the context of multivariate ordinal data without covariates, the latent response distribution was modeled with a DP mixture of multivariate normals. In the absence of covariates, this model is sufficiently flexible to uncover essentially any pattern in a contingency table while 
using fixed cut-offs. This represents a significant advantage relative to the parametric models discussed, 
since the estimation of cut-offs requires nonstandard inferential techniques, such as hybrid Markov 
chain Monte Carlo (MCMC) samplers \citep{johnson} and reparamaterization to achieve transformed 
cut-offs which do not have an order restriction \citep{chen}.

The preceding discussion reflects the fact that there are significant challenges involved in fitting 
multivariate probit models, and a large amount of research has been dedicated to providing new inferential 
techniques in this setting. While the simplicity of its model structure and interpretability of its parameters 
make the probit model appealing to practitioners, the assumptions of linear covariate effects, and 
normality on the latent variables are restrictive. Hence, from both the methodological
and practical point of view, it is important to explore more flexible modeling and inference techniques, especially for the setting with multivariate ordinal responses. Semiparametric methods for binary regression are more common, since in this case there is a single 
regression function to be modeled. When taken to the setting involving a single ordinal response 
with $C\geq 3$ classifications, it becomes much harder to incorporate flexible priors for each of 
the $C-1$ probability response curves. And, semiparametric prior specifications appear 
daunting in the multivariate ordinal regression setting where, in addition to general regression 
relationships, it is desirable to achieve flexible dependence structure between the ordinal responses.

Motivated by these limitations of parametric and semiparametric prior probability models, 
our goal is to develop a Bayesian nonparametric regression model for univariate and multivariate 
ordinal responses, which enables flexible inference for both the conditional response distribution 
and for regression relationships. 
%The way in which the covariates affect the responses is driven by the data, as we do not assume a linear 
%relationship between the latent responses and covariates as in standard probit regression and its variations, 
%or any independence assumptions in the covariate effects.  
We focus on problems where the covariates can be treated as random, and model the joint 
distribution of the covariates and latent responses with a DP mixture of multivariate normals 
to induce flexible regression relationships, as well as general dependence structure in the 
multivariate response distribution. In many fields -- such as biometry, economics, and the 
environmental and social sciences -- the assumption of random covariates is appropriate, 
indeed we argue necessary, as the covariates are not fixed prior to data collection.

In addition to the substantial distributional flexibility, an appealing aspect of the nonparametric 
modeling approach taken is that the cut-offs may be fixed, and the mixture kernel covariance 
matrix left unrestricted. Regarding the latter, we show that all parameters of the normal mixture 
kernel are identifiable provided each ordinal response comprises more than two classifications. 
For the former, we demonstrate that, with fixed cut-offs, our model can approximate arbitrarily 
well any set of probabilities on the ordinal outcomes. To this end, we prove that the induced prior 
on the space of mixed ordinal-continuous distributions assigns positive probability to all 
Kullback-Leibler (KL) neighborhoods of all densities in this space
{that satisfy certain regularity conditions.}

We are primarily interested in regression relationships, and demonstrate that we can obtain inference 
for a variety of nonlinear as well as more standard shapes for ordinal regression curves, using data 
examples chosen from fields for which the methodology is particularly
relevant. The joint response-covariate modeling approach yields more
flexible inference for regression relationships than structured normal
mixture models 
for the response distribution, such as the ones in \citet{chen} and \citet{BaoHanson}. Moreover, as a consequence 
of the joint modeling framework, inferences for the conditional covariate distribution given specific 
ordinal response categories, or inverse inferences, are also available. These relationships may be of 
direct interest in certain applications. Also of interest are the associations between the ordinal variables. 
These are described by the correlations between the latent variables in the ordinal probit model, 
termed polychoric correlations in the social sciences \citep[e.g.,][]{olsson}.  Using a data set of 
ratings assigned to essays by multiple raters, we apply our model to determine regions of the 
covariate space as well as the levels of ratings at which pairs of raters tend to agree or disagree. 
We contrast our approach with the Bayesian methods for studying multirater agreement 
%in the context of ordinal regression described 
in \cite{johnson} and \cite{dalal}.

The work in this paper forms the sequel to \citet{deyoreo}, where we have developed the joint response-covariate modeling approach for binary regression, which requires use of more structured mixture kernels to satisfy necessary identifiability restrictions. The joint response-covariate modeling approach with categorical variables has been explored in a few other papers, including \citet{shahbaba}, \citet{dunson}, 
\citet{hannah}, and \citet{papa:richardson:best}. \citet{shahbaba} considered classification of a univariate 
response using a multinomial logit kernel, and this was extended by \citet{hannah} to accommodate 
alternative response types with mixtures of generalized linear models. \citet{dunson} studied DP mixtures 
of independent kernels, and \citet{papa:richardson:best} build a model for spatially indexed data of mixed type (count, 
categorical, and continuous). However, these models would not be suitable for ordinal data, or, particularly 
in the first three cases, when inferences are to be made on the association between ordinal response variables.

The rest of the article is organized as follows. In Section \ref{sec:method}, we formulate the DP mixture 
model for ordinal regression, including study of the theoretical properties discussed above (with technical 
details given in the Appendix). We discuss prior specification and
posterior inference, and indicate 
%the necessary 
{possible} modifications when binary responses are present. The methodology is applied in Section 
\ref{sec:applications} to ozone concentration data, two data examples from econometrics, and a 
multirater agreement problem. Finally, Section \ref{sec:disc} concludes with a discussion.

%
%-------------  Section 2: Methods ------------------------------------------
%

\section{Methodology}
\label{sec:method}

\subsection{Model formulation}
Suppose that $k$ ordinal categorical variables are recorded for each
of $n$ individuals, along with $p$ continuous covariates, so that for
individual $i$ we observe a response vector
$\boldsymbol{y}_i=(y_{i1},\dots,y_{ik})$ and a covariate vector
$\boldsymbol{x}_i=(x_{i1},\dots,x_{ip})$, with
$y_{ij}\in\{1,\dots,C_j\}$, and $C_j>2$. Introduce latent continuous
random variables $\boldsymbol{z}_i=(z_{i1},\dots,z_{ik})$,
$i=1,\dots,n$, such that $y_{ij}=l$ if and only if
$\gamma_{j,l-1}<z_{ij}\leq\gamma_{j,l}$, for $j=1,\dots,k$, and
$l=1,\dots,C_j$. For example, in a biomedical application, $y_{i1}$
and $y_{i2}$ could represent severity of two different symptoms of
patient $i$, recorded on a categorical scale ranging from ``no
problem'' to ``severe'', along with covariate information weight, age,
and blood pressure. The assumption that the ordinal responses
represent discretized versions of latent continuous responses is
natural for many settings, such as the one considered here.  Note also
that the assumption of random covariates is appropriate in this
application, and that the medical measurements are all related and
arise from some joint stochastic mechanism.  This motivates our focus
on building a model for the joint density
$f(\boldsymbol{z},\boldsymbol{x})$, which is 
a continuous density of dimension $k+p$, which in turn implies a 
model for the conditional response distribution $f(\boldsymbol{y}\mid \boldsymbol{x})$.

To model $f(\boldsymbol{z},\boldsymbol{x})$ in a flexible way, we use
a DP mixture of multivariate normals model, mixing on the mean vector
and covariance matrix. That is, we assume 
$(\boldsymbol{z}_i,\boldsymbol{x}_i)\mid G\stackrel{iid}{\sim}
\int \text{N}(\boldsymbol{z}_i,\boldsymbol{x}_i \mid
\boldsymbol{\mu},\boldsymbol{\Sigma}) \text{d}G(\boldsymbol{\mu},\boldsymbol{\Sigma})$, 
and place a DP prior \citep{ferguson} on the random mixing
distribution $G$. This DP mixture 
%curve fitting approach to regression 
{density regression approach} 
was first proposed by \citet{muller} in the context of
continuous variables, and has been more recently studied under
different modeling scenarios; see, e.g., \citet{mullerquintana},
\citet{parkdunson}, \citet{taddy}, \citet{TaddyKottas2012}, and
\citet{wade}.
{\citet{Norets} established large support and other
theoretical properties for conditional distributions that are induced 
from finite mixtures of multivariate normals, where the response 
and covariates may be either discrete or continuous variables.}

The hierarchical model is formulated by introducing a latent 
mixing parameter $\boldsymbol{\theta_i}=(\boldsymbol{\mu}_i,\boldsymbol{\Sigma}_i)$ for each data vector:
\begin{equation}
\label{eqn:hier_ord}
(\boldsymbol{z}_i,\boldsymbol{x}_i) \mid \boldsymbol{\theta}_i\stackrel{ind.}{\sim}\text{N}(\boldsymbol{\mu}_i,\boldsymbol{\Sigma}_i); \,\,\,\,\,
\boldsymbol{\theta}_i\mid  G \stackrel{iid}{\sim} G, \quad i=1,\dots,n
\end{equation}
where $G \mid \alpha,\boldsymbol{\psi} \sim \text{DP}(\alpha,G_0(\cdot
\mid \boldsymbol{\psi}))$, with base (centering) distribution 
$G_0(\boldsymbol{\mu},\boldsymbol{\Sigma} \mid \boldsymbol{\psi})=$
$\text{N}(\boldsymbol{\mu} \mid \boldsymbol{m},\boldsymbol{V})
\text{IW}(\boldsymbol{\Sigma} \mid \nu,\boldsymbol{S})$. 
The model is completed with hyperpriors on $\boldsymbol{\psi}=$ 
$(\boldsymbol{m},\boldsymbol{V},\boldsymbol{S})$, and a prior on $\alpha$:
\begin{equation}\label{eqn:hyperpriors} \boldsymbol{m}\sim \text{N}(\boldsymbol{a}_{\boldsymbol{m}},\boldsymbol{B}_{\boldsymbol{m}}),\quad \boldsymbol{V}\sim \text{IW}(a_{\boldsymbol{V}},\boldsymbol{B}_{\boldsymbol{V}}),\quad \boldsymbol{S}\sim \text{W}(a_{\boldsymbol{S}},\boldsymbol{B}_{\boldsymbol{S}}),\quad \alpha\sim \text{gamma}(a_\alpha,b_\alpha),\end{equation}
where W$(a_{\boldsymbol{S}},\boldsymbol{B}_{\boldsymbol{S}})$ denotes a Wishart distribution with mean $a_{\boldsymbol{S}}\boldsymbol{B}_{\boldsymbol{S}}$, and IW$(a_{\boldsymbol{V}},\boldsymbol{B}_{\boldsymbol{V}})$ 
denotes an inverse-Wishart distribution with mean $(a_{\boldsymbol{V}}-(k+p)-1)^{-1}\boldsymbol{B}_{\boldsymbol{V}}$.

According to its constructive definition \citep{sethuraman}, the DP generates almost surely discrete 
distributions, $G=\sum_{l=1}^{\infty}p_l \delta_{\boldsymbol{\vartheta}_l}$. 
The atoms ${\boldsymbol{\vartheta}}_l$ are independent realizations from $G_0$, and the weights are determined through stick-breaking from beta distributed random variables with parameters 1 and $\alpha$. That is, $p_1=v_1$, and $p_l=v_l\prod_{r=1}^{l-1}(1-v_r)$, for $l=2,3,\dots$, with $v_l\stackrel{iid}{\sim}\text{beta}(1,\alpha)$. The discreteness of the DP allows for ties in the $\boldsymbol{\theta}_i$, so that in practice less than $n$ distinct values for the $\{\boldsymbol{\theta}_i\}$ are imputed. The data is therefore clustered into a typically small number of groups relative to $n$, with the number of clusters $n^*$ controlled by parameter $\alpha$, 
where larger values favor more clusters \citep{escwest}. From the constructive definition for $G$, the prior model for $f(\boldsymbol{z},\boldsymbol{x})$ has an almost sure representation as a countable mixture of multivariate normals, and the proposed model can therefore be viewed as a nonparametric extension of the multivariate probit model, albeit with random covariates.

This implies a countable mixture of normal distributions (with covariate-dependent weights) for 
$f(\boldsymbol{z}\mid  \boldsymbol{x},G)$,
from which the latent $\boldsymbol{z}$ may be integrated out to reveal the induced model for the ordinal regression relationships.  In general, for a multivariate response $\boldsymbol{Y}=(Y_1,\dots,Y_k)$ with an associated covariate vector $\boldsymbol{X}$, the probability that $\boldsymbol{Y}$ takes on the values $\boldsymbol{l}=(l_1,\dots,l_k)$, where $l_j\in \{1,\dots,C_j\}$, for $j=1,\dots,k$, can be expressed as
\begin{equation}\label{eqn:regfull}
\mathrm{Pr}(\boldsymbol{Y}=\boldsymbol{l}\mid  \boldsymbol{x},G) =
\sum_{r=1}^{\infty}w_r(\boldsymbol{x})\int_{\gamma_{k,l_k-1}}^{\gamma_{k,l_k}}\cdot\cdot\cdot 
\int_{\gamma_{1,l_1-1}}^{\gamma_{1,l_1}}\text{N}(\boldsymbol{z} \mid \boldsymbol{m}_r(\boldsymbol{x}),\boldsymbol{S}_r)
\text{d}\boldsymbol{z}
\end{equation}
with covariate-dependent weights 
$w_r(\boldsymbol{x})\propto p_r \text{N}(\boldsymbol{x} \mid \boldsymbol{\mu}_r^{x},\boldsymbol{\Sigma}_r^{xx})$, and mean vectors $\boldsymbol{m}_r(\boldsymbol{x})=\boldsymbol{\mu}_r^{z}+\boldsymbol{\Sigma}_r^{zx}(\boldsymbol{\Sigma}_r^{xx})^{-1}(\boldsymbol{x}-\boldsymbol{\mu}_r^{x})$, and covariance matrices $\boldsymbol{S}_r=\boldsymbol{\Sigma}_r^{zz}-\boldsymbol{\Sigma}_r^{zx}(\boldsymbol{\Sigma}_r^{xx})^{-1}\boldsymbol{\Sigma}_r^{xz}$. Here, $(\boldsymbol{\mu}_r,\boldsymbol{\Sigma}_r)$ are the atoms in the DP prior constructive definition, where $\boldsymbol{\mu}_r$ is partitioned into $\boldsymbol{\mu}_r^{z}$ and $\boldsymbol{\mu}_r^{x}$ according to random vectors $\boldsymbol{Z}$ and $\boldsymbol{X}$, and $(\boldsymbol{\Sigma}_r^{zz},\boldsymbol{\Sigma}_r^{xx},\boldsymbol{\Sigma}_r^{zx}, \boldsymbol{\Sigma}_r^{xz})$ are the components of the corresponding partition of covariance matrix $\boldsymbol{\Sigma}_r$.

To illustrate, consider a bivariate response $\boldsymbol{Y}=(Y_1,Y_2)$, with covariates $\boldsymbol{X}$. The probability assigned to the event $(Y_1=l_1)\cap (Y_2=l_2)$ is obtained using (\ref{eqn:regfull}), which involves evaluating bivariate normal distribution functions.  However, one may be interested in the marginal relationships between individual components of $\boldsymbol{Y}$ and the covariates. Referring to the example given at the start of this section, we may obtain the probability that both symptoms are severe as a function of $\boldsymbol{X}$, but also how the first varies as a function of $\boldsymbol{X}$. The marginal inference, $\mathrm{Pr}(Y_1=l_1\mid  \boldsymbol{x},G)$, is given by the expression
\begin{equation}\label{eqn:marg}
\sum_{r=1}^{\infty}w_r(\boldsymbol{x})\left\{\Phi \left(\frac{\gamma_{1,l_1}-m_r(\boldsymbol{x})}{s_r^{1/2}}\right)-\Phi \left(\frac{\gamma_{1,l_1-1}-m_r(\boldsymbol{x})}{s_r^{1/2}}\right)\right\}
\end{equation}
where $m_r(\boldsymbol{x})$ and $s_r$ are the conditional mean and
variance for $z_1$ conditional on $\boldsymbol{x}$ implied by the joint distribution 
$\mathrm{N}(\boldsymbol{z},\boldsymbol{x} \mid \boldsymbol{\mu}_r,\boldsymbol{\Sigma}_r)$. 
Expression (\ref{eqn:marg}) provides also the form of the ordinal regression curves in the case of a single response.

Hence, the implied regression relationships have a mixture structure with component-specific kernels 
which take the form of parametric probit regressions, and weights which are covariate-dependent. 
This structure enables inference for non-linear response curves, by favoring a set of 
parametric models with varying probabilities depending on the location in the covariate space. 
The limitations of parametric probit models -- including relative covariate effects which 
are constant in terms of the covariate and the ordinal level, monotonicity, and the single-crossing 
property of the response curves -- are thereby overcome.

\subsection{Model properties}
\label{sec:ord_model_prop}

In (\ref{eqn:hier_ord}), $\boldsymbol{\Sigma}$ was left an unrestricted covariance matrix.
%and assigned an inverse-Wishart base distribution in $G_0$. 
As an alternative to working with correlation matrices under the probit model, identifiability 
can be achieved by fixing $\gamma_{j,2}$ (in addition to $\gamma_{j,1}$), for $j=1,\dots,k$. 
As shown here, analogous results can be obtained in the random covariate setting. In particular, 
Lemma 1 establishes identifiability for $\boldsymbol{\Sigma}$ as a general covariance matrix {in a parametric ordered probit model with random covariates}, 
assuming all cut-offs, $\gamma_{j,1},\dots,\gamma_{j,C_j-1}$, for $j=1,\dots,k$, are fixed. Here, 
model identifiability refers to likelihood identifiability of the mixture kernel parameters in the 
induced model for $(\boldsymbol{Y},\boldsymbol{X})$ {under a single mixture component.}

\vspace{0.15cm}

\noindent
{\bf Lemma 1}. \emph{ {Consider a multivariate normal model, $\text{N}(\boldsymbol{Z},\boldsymbol{X}\mid\boldsymbol{\mu},\boldsymbol{\Sigma})$, for the joint distribution of latent responses $\boldsymbol{Z}$ and covariates $\boldsymbol{X}$, with fixed cut-off points defining the ordinal responses $\boldsymbol{Y}$ through $\boldsymbol{Z}$.
Then, the parameters $\boldsymbol{\mu}$ and $\boldsymbol{\Sigma}$ are identifiable in the 
implied model for $(\boldsymbol{Y},\boldsymbol{X})$, and therefore in the kernel of the 
implied mixture model for $(\boldsymbol{Y},\boldsymbol{X})$, provided $C_j>2$, for $j=1,\dots,k$.}}

\vspace{0.1cm}

Refer to Appendix A for a proof of this result. If $C_j=2$ for some $j$, additional restrictions are needed 
for identifiability, as discussed in Section \ref{sec:bin}.

{We emphasize that Lemma 1 does not imply that the 
parameters of the mixture model are identifiable, rather it
establishes identifiability for the kernel of the mixture model. 
Identifiability is a basic model property for parametric probit models 
\citep{hobert,mcculloch,koop,talhouk}, and is achieved for the 
special one-component case of our model, a multivariate ordered 
probit model with random covariates, by fixing the cut-offs.}
However, 
this may appear to be a significant restriction, as under the parametric probit setting, fixing all 
cut-off points would prohibit the model from being able to adequately assign probabilities to 
the regions determined by the cut-offs. We therefore seek to determine if the nonparametric 
model with fixed cut-offs is sufficiently flexible to accommodate 
{general distributions} 
for $(\boldsymbol{Y},\boldsymbol{X})$ and also for $\boldsymbol{Y} \mid \boldsymbol{X}$.
\cite{kottas} provide an informal argument 
that the normal DP mixture model for multivariate ordinal responses (without covariates) can approximate 
arbitrarily well any probability distribution for a contingency table. The basis for this argument is that, in 
the limit, one mixture component can be placed within each set of cut-offs corresponding to a specific 
ordinal vector, with the mixture weights assigned accordingly to each cell.

Here, we provide a proof of the full support of our more general model for ordinal-continuous distributions. 
%A prior model has large support if it can generate densities which are arbitrarily close to any true 
%data-generating density.  
In addition to being a desirable property on its own, the ramifications of full support for the prior
model are significant, as it is a key condition for the study of posterior consistency 
\citep[e.g.,][]{ghosh}. Using the KL divergence to define density neighborhoods, a particular density 
$f_0$ is in the KL support of the prior $\mathcal{P}$, if 
$\mathcal{P}(K_\epsilon(f_0)) >0$, for any $\epsilon>0$, where $K_\epsilon(f_0)=$
$\{f:\int f_0(\boldsymbol{w})\log(f_0(\boldsymbol{w})/f(\boldsymbol{w})) \text{d}\boldsymbol{w}<\epsilon\}$. 
The KL property is satisfied if any density
{(typically subject to regularity conditions)}
in the space of interest is in the KL support of the prior.

It has been established that the DP location normal mixture prior
model satisfies the KL property \citep{wu}. 
That is, if the mixing distribution $G$ is assigned a DP prior on the space of random distributions for 
$\boldsymbol{\mu}$, and a normal kernel is chosen such that
$f(\boldsymbol{w} \mid G,\boldsymbol{\Sigma})=$
$\int \mathrm{N}(\boldsymbol{w} \mid \boldsymbol{\mu},\boldsymbol{\Sigma})dG(\boldsymbol{\mu})$, with 
$\boldsymbol{\Sigma}$ a diagonal matrix, then the induced prior model for densities has full KL support.
Letting this induced prior be denoted by $\mathcal{P}$, and modeling the joint distribution of 
$(\boldsymbol{X},\boldsymbol{Z})$ with a DP location mixture of normals, the KL property yields:
\begin{equation}
\label{eqn:KLprop}
\mathcal{P}\left(\{f: \int  f_0(\boldsymbol{x},\boldsymbol{z})
\log(f_0(\boldsymbol{x},\boldsymbol{z})/f(\boldsymbol{x},\boldsymbol{z}))
\text{d}\boldsymbol{x} \text{d}\boldsymbol{z} <\epsilon \} \right)>0,
\,\,\,\,\, \text{for any} \,\,\, \epsilon>0
\end{equation}
for all densities $f_0(\boldsymbol{x},\boldsymbol{z})$ on $\mathbb{R}^{p+k}$
{that satisfy certain regularity conditions.}

To establish the KL property of the prior on mixed ordinal-continuous distributions for 
$(\boldsymbol{X},\boldsymbol{Y})$ induced from multivariate continuous distributions for 
$(\boldsymbol{X},\boldsymbol{Z})$, consider a generic data-generating distribution $p_0(\boldsymbol{x},\boldsymbol{y})\in\mathcal{D^*}$. Here, $\mathcal{D^*}$ denotes the 
space of distributions on $\mathbb{R}^p\times \{1,\dots,C_1\}\times
\cdot\cdot\cdot \times \{1,\dots,C_k\}$.  
Let $f_0(\boldsymbol{x},\boldsymbol{z}) \in \mathcal{D}$, where $\mathcal{D}$ 
denotes the space of densities on $\mathbb{R}^{p+k}$, such that
\begin{equation}
\label{eqn:mapping}
p_0(\boldsymbol{x},l_1,\dots,l_k) = \int_{\gamma_{k,l_k-1}}^{\gamma_{k,l_k}} \cdot\cdot\cdot \int_{\gamma_{1,l_1-1}}^{\gamma_{1,l_1}}f_0(\boldsymbol{x},z_1,\dots,z_k) \, \text{d}z_{1} \cdot\cdot\cdot \text{d}z_{k}
\end{equation}
for any $(l_{1},...,l_{k})$, where $l_j\in\{1,\dots,C_j\}$, for $j=1,\dots,k$.  
That is, $f_0(\boldsymbol{x},\boldsymbol{z})$ is 
{the density of a latent continuous distribution}
which induces the corresponding distribution on the ordinal responses.  
Note that at least one $f_0\in \mathcal{D}$ exists for each $p_0\in \mathcal{D^*}$, with one such $f_0$ 
described in Appendix B2.

The next lemma (proved in Appendix B1) establishes that, as a consequence 
of the KL property of the DP mixture of normals (\ref{eqn:KLprop}), the prior assigns positive probability 
to all KL neighborhoods of $p_0(\boldsymbol{x},\boldsymbol{y})$, as well as to all KL neighborhoods 
of the implied conditional distributions $p_0(\boldsymbol{y} \mid \boldsymbol{x})$. 
{We prove the lemma assuming that $p_{0}$
satisfies generic regularity conditions such that the
corresponding $f_{0}$ is in the KL support of the prior. Hence, the
result may be useful for other prior model settings for the latent
continuous distribution $f(\boldsymbol{z},\boldsymbol{x})$. 
Lemma 2 extends to the regression setting analogous KL 
support results for discrete and mixed 
discrete-continuous distributions \citep{canale,canale2015}. 
To ensure the KL property is satisfied for the proposed model, we also include
in Appendix B2 a set of conditions for $p_{0}$ under which a specific
corresponding $f_{0}$ satisfies (\ref{eqn:KLprop}), in particular, it satisfies 
the general conditions for the KL property given in Theorem 2 of \cite{wu}.}

\vspace{0.15cm}

\noindent
{\bf Lemma 2}. 
\emph{{Assume that the distribution of a mixed ordinal-continuous 
random variable, $p_0(\boldsymbol{x},\boldsymbol{y})\in \mathcal{D^*}$, 
satisfies appropriate regularity conditions such that for the
corresponding density, $f_0(\boldsymbol{x},\boldsymbol{z})\in \mathcal{D}$,
we have} $\mathcal{P}(K_\epsilon(f_0(\boldsymbol{x},\boldsymbol{z})))>0$, 
for any $\epsilon>0$. Then, 
$\mathcal{P}(K_\epsilon(p_0(\boldsymbol{x},\boldsymbol{y}))) > 0$ and 
$\mathcal{P}(K_{\epsilon}(p_0(\boldsymbol{y}\mid  \boldsymbol{x}))) > 0$, for any $\epsilon>0$.} 

\vspace{0.1cm}

Lemma 2 establishes full KL support for a model arising from a DP location normal mixture, 
a simpler version of our model. 
%Combined together, the properties 
%of identifiability and full support reflect a major advantage of the proposed model. That is, it 
%can approximate arbitrarily well any distribution on $(\boldsymbol{Y},\boldsymbol{X})$, as well as any 
%conditional distribution for $(\boldsymbol{Y}\mid  \boldsymbol{X})$, while at the same time avoiding 
%the need to impute cut-off points or work with correlation matrices, both of which are major challenges 
%in fitting multivariate probit models.  
{The KL property reflects a major advantage of the
proposed modeling approach in that it offers flexible inference for multivariate 
ordinal regression while avoiding the need to impute cut-off points, 
which is a major challenge in fitting multivariate probit models.}

The cut-off points can be fixed to arbitrary increasing values, which, for
instance, may be equally spaced and centered at zero. As confirmed 
empirically with simulated data (refer to Section \ref{sec:applications} for details), 
the choice of cut-offs does not affect inferences for the ordinal regression 
relationships, only the center and scale of the latent variables,
which must be interpreted relative to the cut-offs. 
%
%The cut-offs must be specified initially, as they play a role in the priors since they determine the location and scale of %$\boldsymbol{Z}$. One strategy for specifying the cut-offs is to set $\gamma_{j,i}=-\gamma_{j,J-i}$, for $j=1,\dots,k$, and %$i=1,\dots,(C_j-1)/2$, if $C_j$ is odd, such that the cut-offs are equally spaced from $\gamma_{j,1}=-a$ to $\gamma_{j,J-1}=a$, for %some positive constant $a$.  If $C_j$ is even, then set $\gamma_{j,C_j/2}=0$, and $\gamma_{j,i}=-\gamma_{j,J-i}=a$, for %$j=1,\dots,C_j$, and $i=1,\dots,C_j/2-1$. The cut-offs for each latent variable $Z_j$ are therefore centered at zero, and their scale %(or the choice of the constant $a$) determines the range of the latent responses, and thus all inference will be interpreted relative to %these cut-offs. 
%

\subsection{Prior specification}
\label{sec:prior}

To implement the model, we need to specify the parameters of the hyperpriors in (\ref{eqn:hyperpriors}).  
A default specification strategy is developed by considering the limiting case of the model as 
$\alpha \rightarrow 0^+$, which results in a single normal distribution for $(\boldsymbol{Z},\boldsymbol{X})$.  
This limiting model is essentially the  multivariate probit model, with the addition of random covariates.  
The only covariate information we use here is an approximate center and range for each covariate, 
denoted by $\boldsymbol{c}^{x}=$ $(c_1^x,\dots,c_p^x)$ and $\boldsymbol{r}^{x}=$ $(r_1^x,\dots,r_p^x)$.  
Then $c_m^{x}$ and $r_m^{x}/4$ are used as proxies for the marginal mean and standard deviation of $X_m$. 
We also seek to center and scale the latent variables appropriately, using the cut-offs. Since $Y_j$ 
is supported on $\{1,\dots,C_j\}$, latent continuous variable $Z_j$ must be supported on values 
slightly below $\gamma_{j,1}$, up to slightly above $\gamma_{j,C_j-1}$. We therefore use $r_j^z/4$, 
where $r_j^{z}=(\gamma_{j,C_j-1}-\gamma_{j,1})$, as a proxy for the standard deviation of $Z_j$.

Under $(\boldsymbol{Z},\boldsymbol{X})\mid \boldsymbol{\mu},\boldsymbol{\Sigma}\sim \text{N}(\boldsymbol{\mu},\boldsymbol{\Sigma})$, we have $\text{E}(\boldsymbol{Z},\boldsymbol{X})=$
$\boldsymbol{a_m}$, and $\text{Cov}(\boldsymbol{Z},\boldsymbol{X})=$
\linebreak
$a_{\boldsymbol{S}}\boldsymbol{B}_{\boldsymbol{S}}(\nu-d-1)^{-1}$ + $\boldsymbol{B}_{\boldsymbol{V}}(a_{\boldsymbol{V}}-d-1)^{-1}$
+ $\boldsymbol{B}_{\boldsymbol{m}}$, with $d=p+k$. Then, assuming each set of cut-offs 
$(\gamma_{j,0},\dots,\gamma_{j,C_j})$ are centered at 0, we fix $a_m=(0,\dots,0,\boldsymbol{c}^{x})$. 
Letting $\boldsymbol{D}=$
$\text{diag}\{(r_1^{{z}}/4)^2,\dots,(r_k^{{z}}/4)^2,(r_1^{{x}}/4)^2,\dots,(r_p^{{x}}/4)^2\}$, each of the three terms in $\text{Cov}(\boldsymbol{Z},\boldsymbol{X})$ can be assigned an equal proportion of the total covariance, and set to $(1/3)\boldsymbol{D}$, or to $(1/2)\boldsymbol{D}$ to inflate the variance slightly.  
For dispersed but proper priors with finite expectation, $\nu$, $a_{\boldsymbol{V}}$, and $a_{\boldsymbol{S}}$ can 
be fixed to $d+2$. Fixing these parameters allows for $\boldsymbol{B_S}$ and $\boldsymbol{B_V}$ to be 
determined accordingly, completing the default specification strategy for the hyperpriors of 
$\boldsymbol{m}$, $\boldsymbol{V}$, and $\boldsymbol{S}$.

%
%Although we have developed an approach to prior specification which utilizes the model 
%for $(\boldsymbol{Z},\boldsymbol{X})$, the focus of this work is in modeling regression functions, 
%so we should also consider the priors which are induced for the regression relationships. 
In the strategy outlined above, the form of $\text{Cov}(\boldsymbol{Z},\boldsymbol{X})$ was diagonal, 
such that a priori we favor independence between $\boldsymbol{Z}$ and $\boldsymbol{X}$ within a 
particular mixture component. Combined with the other aspects of the prior specification approach,
this generally leads to prior means for the ordinal regression curves which do not have any trend across 
the covariate space. Moreover, the corresponding prior uncertainty bands span a significant portion
of the unit interval.
%In the expressions for the regression functions in (\ref{eqn:regfull}) and (\ref{eqn:marg}), it is easy to see that if %$\boldsymbol{\Sigma}^{zx}_l=\boldsymbol{0}$ for all $l$, then $\boldsymbol{m}_l(\boldsymbol{x})=\boldsymbol{\mu}^{z}_l$, %and the probabilities no longer depend on $\boldsymbol{x}$. This leads to regression curves which are flat in the prior mean, and %not increasing or decreasing over the covariate space, and this method can therefore be considered if noninformative priors are %desired, or when it is unknown how the response variables vary over $\boldsymbol{X}$.
%

\subsection{Posterior inference}
\label{sec:post}

{There exist many MCMC algorithms for DP mixtures, including
  marginal samplers \citep[e.g.,][]{neal}, slice sampling algorithms
  \citep[e.g.,][]{kalli}, and methods that involve
  truncation of the mixing distribution $G$
  \citep[e.g.,][]{ishjames}. If the inference objectives are limited
  to posterior predictive estimation, which does not require posterior
  samples for $G$, the former two options may be preferable
  since they avoid truncation of $G$. However, under the joint
  modeling approach for the response-covariate distribution, posterior
  simulation must be extended to $G$ to obtain full inference for
  regression functionals, and this requires truncation regardless of
  the chosen MCMC approach. Since our contribution is on flexible mixture
  modeling for ordinal regression,  we view the choice of the 
posterior simulation method as one that should balance efficiency with ease of implementation.}

The MCMC posterior simulation method we utilize is based on a finite truncation approximation to 
the countable mixing distribution $G$, using the DP stick-breaking representation. The blocked 
Gibbs sampler \citep{ishwaran,ishjames} replaces the countable sum with a finite sum, $G_N=$ 
$\sum_{l=1}^{N} p_{l} \delta_{(\boldsymbol{\mu}_l,\boldsymbol{\Sigma}_l)}$, with 
$(\boldsymbol{\mu}_l,\boldsymbol{\Sigma}_l)$ i.i.d. from $G_{0}$, for $l=1,...,N$.
Here, the first $N-1$ elements of $\boldsymbol{p}=$ $(p_{1},...,p_{N})$ are equivalent to those 
in the countable representation of $G$, whereas $p_{N}=$ $1 - \sum_{l=1}^{N-1} p_{l}$. 
Under this approach, the posterior samples for model parameters yield posterior samples 
for $G_N$, and therefore full inference is available for mixture functionals. 

{The truncation level $N$ can be chosen to any desired level of
accuracy, using standard DP properties. In particular, the expectation 
for the partial sum of the original DP weights, 
$\mathrm{E}(\sum\nolimits_{l=1}^{N} p_{l} \mid \alpha)=$
$1 - \{ \alpha/(\alpha + 1) \}^{N}$, can be averaged over the prior for $\alpha$
to estimate the marginal prior expectation,
$\mathrm{E}(\sum\nolimits_{l=1}^{N} p_{l})$, which is then used to specify 
$N$ given a tolerance level for the approximation.
For instance, for the data example of Section \ref{sec:ozone}, we set
$N=50$, which under a $\mathrm{gamma}(0.5,0.5)$ prior for $\alpha$, yields 
$\mathrm{E}(\sum\nolimits_{l=1}^{50} p_{l}) \approx 0.99994$.} 
%
%An alternative approach, which involves also the sample size, is
%available through Theorem 2 in \citet{ishjames} which provides an
%(approximate) upper bound of $4 n \exp\{ -(N-1)/\alpha \}$ on the 
%$L_{1}$ distance between the prior predictive probability of the sample 
%under the infinite dimensional prior $G$ and its truncated version 
%$G_{N}$.}
%

To express the hierarchical model for the data after replacing $G$ with $G_N$, introduce configuration 
variables $(L_1,\dots,L_n)$, such that $L_i=l$ if and only if
$\boldsymbol{\theta}_i=$ $(\boldsymbol{\mu}_l,\boldsymbol{\Sigma}_l)$,
for $i=1,...,n$ and $l=1,...,N$. Then, the model for the data becomes
\begin{eqnarray*}
 y_{ij}=l \quad \mathrm{iff} \quad \gamma_{j,l-1}<z_{ij}\leq\gamma_{j,l},\quad i=1,\dots,n,\quad j=1,\dots,k
 \nonumber \\
(\boldsymbol{z}_i,\boldsymbol{x}_i) \mid \{ (\boldsymbol{\mu}_l,\boldsymbol{\Sigma}_l): l=1,...,N \},L_i
\stackrel{ind.}{\sim}\text{N}(\boldsymbol{\mu}_{L_i},\boldsymbol{\Sigma}_{L_i}),\quad i=1,\dots,n
\nonumber \\
L_{i} \mid  \boldsymbol{p} \stackrel{iid}{\sim} \sum\nolimits_{l=1}^{N}p_{l}\delta_{l}(L_{i}), \quad i=1,...,n
\nonumber \\
(\boldsymbol{\mu}_l,\boldsymbol{\Sigma}_l)\mid  \boldsymbol{\psi} 
\stackrel{iid}{\sim}\text{N}(\boldsymbol{\mu}_l \mid \boldsymbol{m},\boldsymbol{V})
\text{IW}(\boldsymbol{\Sigma}_l \mid \nu,\boldsymbol{S}),\quad l=1,\dots,N
\nonumber 
\end{eqnarray*}
where the prior density for $\boldsymbol{p}$ is given by 
$\alpha^{N-1} p_{N}^{\alpha-1} (1-p_1)^{-1} (1-(p_1+p_2))^{-1} \times \dots \times (1 - \sum\nolimits_{l=1}^{N-2}p_l)^{-1}$, 
which is a special case of the generalized Dirichlet distribution. The full model is completed with the 
conditionally conjugate priors on $\boldsymbol{\psi}$ and $\alpha$ as given in (\ref{eqn:hyperpriors}).

All full posterior conditional distributions are readily sampled, enabling efficient
Gibbs sampling from the joint posterior distribution of the model above. Conditional on the latent
responses $\boldsymbol{z}_{i}$, we have standard updates for the parameters of a normal DP mixture
model. And conditional on the mixture model parameters, each $z_{ij}$, for $i=1,\dots,n$ and 
$j=1,\dots,k$, has a truncated normal full conditional distribution supported on the interval 
$(\gamma_{j,y_{ij}-1},\gamma_{j,y_{ij}}]$.
%
%allowing a Gibbs sampler to be used for sampling from the full posterior distribution $p(\boldsymbol{\mu}, %\boldsymbol{\Sigma},\boldsymbol{L},\boldsymbol{p},\alpha,\boldsymbol{\psi},\boldsymbol{z}\mid  \mathrm{data})$. The full %conditional distribution for each $\boldsymbol{\mu}_l$ is normal, that for $\boldsymbol{\Sigma}_l$ is inverse-Wishart, and each %$L_i$ is drawn from the discrete distribution on $\{1,\dots,N\}$.  Each latent $z_{ij}$, $i=1,\dots,n$, $j=1,\dots,k$, has a 
%truncated normal full conditional distribution, supported on the interval $(\gamma_{j,y_{ij}-1},\gamma_{j,y_{ij}}]$.
%

The regression functional $\mathrm{Pr}(\boldsymbol{Y}=\boldsymbol{l}\mid  \boldsymbol{x},G)$ (estimated 
by the truncated version of (\ref{eqn:regfull}) implied by $G_N$) can be computed over a grid in $\boldsymbol{x}$ 
at every MCMC iteration. This yields an entire set of samples for ordinal response probabilities at any covariate 
value $\boldsymbol{x}$ (note that $\boldsymbol{x}$ may include just a portion of the covariate vector or a 
single covariate). As indicated in (\ref{eqn:marg}), in the multivariate setting, we may wish to report
inference for individual components of $\boldsymbol{Y}$ over the covariate space.

In some applications, in addition to modeling how $\boldsymbol{Y}$ varies across $\boldsymbol{X}$, we may 
also be interested in how the distribution of $\boldsymbol{X}$ changes at different ordinal values of 
$\boldsymbol{Y}$. As a feature of the modeling approach, we can obtain inference for 
$f(\boldsymbol{x}\mid  \boldsymbol{y},G)$, for any configuration of ordinal response levels $\boldsymbol{y}$. 
We refer to these inferences as inverse relationships, and illustrate them with the data example of 
Section \ref{sec:credit}.

Under the multivariate response setting, the association between the ordinal variables may also be a key 
target of inference. In the social sciences, the correlations between pairs of latent responses, 
$\text{corr}(Z_r,Z_s)$, are termed polychoric correlations \citep{olsson} when a single multivariate normal 
distribution is used for the latent responses. Under our mixture modeling framework, we can sample a 
single $\text{corr}(Z_r,Z_s)$ at each MCMC iteration according to the corresponding $\boldsymbol{p}$, 
providing posterior predictive inference to assess overall agreement between pairs 
of response variables. As an alternative, and arguably more informative measure of association, we can 
obtain inference for probability of agreement over each covariate, or probability of agreement at each 
ordinal level. These inferences can be used to identify parts of the covariate space where there is 
agreement between response variables, as well as the ordinal values which are associated with higher
levels of agreement. In the social sciences it is common to assess agreement among multiple raters 
or judges who are assigning a grade to the same item. In Section \ref{sec:multirater}, we illustrate our 
methods with such a multirater agreement data example, in which both estimating regression 
relationships and modeling agreement are major objectives.

\subsection{Accommodating binary responses}
\label{sec:bin}

Our methodology focuses on multivariate ordinal responses with $C_j>2$, for all $j$. However, if one 
or more responses is binary, then the full covariance matrix of the normal mixture kernel for 
$(\boldsymbol{Z},\boldsymbol{X})$ is not identifiable. In the univariate probit model, it is standard 
practice to assume that $Z \mid \boldsymbol{\beta} \sim$ 
$\mathrm{N}(\boldsymbol{x}^T\boldsymbol{\beta},1)$, that is, identifiability is achieved by fixing 
$\Sigma^{zz}$. \cite{deyoreo} {show that a similar
  restriction can be used to facilitate identifiability in the binary
  probit model with random covariates, which is built from a
  multivariate normal distribution for the single latent response and covariates.}

%
%When multiple ordinal responses exist and one or more is binary, it follows from the univariate case that we can not hope to %estimate all elements of $\boldsymbol{\Sigma}$, in particular the covariance elements corresponding to the binary responses. %Identifiability in this setting can be accomplished by fixing the diagonal elements of $\boldsymbol{\Sigma}^{zz}$ which %represent the variances of the latent binary responses. The covariance elements $\Sigma^{z_iz_j}$, $i\neq j$, all remain free, %which is important since the association between the responses may be of interest. 
%
Under the general setting involving some binary and some ordinal responses, extending the argument 
that led to Lemma 1, it can be shown that identifiability is accomplished by fixing the diagonal 
elements of $\boldsymbol{\Sigma}^{zz}$ that correspond to the variances of the latent binary responses
(the associated covariance elements remain identifiable).
%The associated covariance elements are identifiable which is important since the association 
%between the responses may be of interest. 
To incorporate this restriction, the inverse Wishart distribution for the component of $G_{0}$ that 
corresponds to $\boldsymbol{\Sigma}$ must be replaced with a more structured specification. 
To this end, a square-root-free Cholesky decomposition of $\boldsymbol{\Sigma}$ \citep{daniels,webb} 
can be used to fix $\Sigma^{zz}$ in the setting with a single binary response \citep{deyoreo}, and this 
is also useful in the multivariate setting. This decomposition expresses $\boldsymbol{\Sigma}$ in terms 
of a unit lower triangular matrix $\boldsymbol{B}$, and a diagonal matrix $\boldsymbol{\Delta}$, with 
elements $\delta_{i} > 0$, $i=1,...,k+p$, such that $\boldsymbol{\Sigma}=$ 
$\boldsymbol{B}^{-1}\boldsymbol{\Delta} (\boldsymbol{B}^{-1})^{T}$. 
The key result is that if $(W_1,\dots,W_r)\sim$
$\mathrm{N}(\boldsymbol{\mu},\boldsymbol{B}^{-1}\boldsymbol{\Delta} (\boldsymbol{B}^{-1})^{T})$, 
then this implies $\text{Var}(W_i\mid  W_{1},\dots,W_{i-1})=$ $\delta_i$, for $i=2,\dots,r$. 
%This was used by \cite{webb} for modeling multivariate binary data. 
Therefore, if $(\boldsymbol{Z},\boldsymbol{X})\sim$
$\mathrm{N}(\boldsymbol{\mu},\boldsymbol{B}^{-1} \boldsymbol{\Delta} (\boldsymbol{B}^{-1})^{T})$, 
with $(Z_1,\dots,Z_r)$ binary, and $(Z_{r+1},\dots,Z_k)$ ordinal, then fixing $\delta_1$ fixes 
$\text{Var}(Z_1)$, fixing $\delta_2$ fixes $\text{Var}(Z_2\mid  Z_1)$, and so on. The scale of the 
latent binary responses may therefore be constrained by fixing $\delta_1$, the variance of the first 
latent binary response, $Z_1$, and the conditional variances $(\delta_2,\dots,\delta_r)$ of the remaining 
latent binary responses $(Z_2,\dots,Z_r)$. The conditional variances $(\delta_{r+1},\dots,\delta_{k+p})$ 
are not restricted, since they correspond to the scale of latent ordinal responses or covariates, 
which are identifiable under our model with fixed cut-offs.

{The restricted version of the model for binary
  responses may be appealing because it implies that the special case of
  the mixture model consisting of a single mixture component (the
  parametric probit model) is identifiable. There is some empirical
  evidence from the literature that forcing the mixture kernel to be identifiable
  can improve MCMC performance \citep[e.g.,][]{dilucca,deyoreo}.
To establish full support for the restricted prior model, the result
of Lemma 2 would need to be extended, since the prior mixture model 
for the latent continuous distribution is modified. 
Alternatively, with sufficient care in MCMC implementation,
the mixture model with the unrestricted kernel covariance matrix 
can be applied with binary responses. Note that Lemma 2 does not
require the $C_j>2$ assumption, that is, it is applicable to settings
with a combination of binary and more general ordinal responses.}

%
%---------------------- Section 3: Data examples ---------------------------------
%

\section{Data Examples}
\label{sec:applications}

The model was extensively tested on simulated data, where the primary goal was to assess how 
well it can estimate regression functionals which exhibit highly non-linear trends. We also 
explored effects of sample size, choice of cut-offs, and number of response categories.
{Some results from simulation studies are presented in the supplementary material.}

The effect of sample size was observed in the uncertainty bands for the regression functions, 
which were reduced in width and became smoother with a larger sample size. The regression 
estimates captured the truth well in all cases, but were smoother and more accurate with more 
data, as expected. Even with a five-category ordinal response and only $200$ observations, 
the model was able to capture quite well non-standard regression curves. 

As discussed previously, the cut-offs may be fixed to arbitrary increasing values, with the 
choice expected to have no impact on inference involving the relationship between 
$\boldsymbol{Y}$ and $\boldsymbol{X}$. To test this point, the model was fit to a synthetic 
data set containing an ordinal response with three categories, using cut-off points of 
$(-\infty,-20,20,\infty)$ and $(-\infty,-5,5,\infty)$, as well as cut-offs $(-\infty,0,0.1,\infty)$, 
which correspond to a narrow range of latent $Z$ values producing response value $Y=2$. 
The ordinal regression function estimates were unaffected by the change in cut-offs. 
The last set of cut-offs forces the model to generate components with small variance (lying in 
the interval $(0,0.1)$), but the resulting regression estimates are unchanged from the previous ones. 

In the data illustrations that follow, the default prior specification strategy outlined in 
Section \ref{sec:prior} was used. The posterior distributions for each component of 
$\boldsymbol{m}$ were always very peaked compared to the prior. 
%indicating the prior on $\boldsymbol{m}$ to be sufficiently diffuse. 
Some sensitivity to the priors was found in terms of posterior learning for hyperparameters 
$\boldsymbol{V}$ and $\boldsymbol{S}$, however this was not reflected in the posterior 
inferences for the regression functions, which displayed little to no change when the priors 
were altered. Regarding the DP precision parameter $\alpha$, we noticed a moderate 
amount of learning for larger data sets, and a small amount for smaller data 
sets, which is consistent with what has been empirically observed about $\alpha$ in DP 
mixtures. The prior for $\alpha$ was in all cases chosen to favor reasonably large 
values, relative to the sample size, for the number of distinct mixture components.

{In all data examples, a total of at least 100,000 MCMC 
samples were taken. The first 5,000 to 10,000 were discarded as burn
in, and the remaining draws were thinned to reduce autocorrelation. 
The models were programmed in R and run on a desktop computer with
Intel 3.4 GHz. To provide an idea of how long the algorithm takes to
run, for the ozone data example, it took approximately 48 seconds (85
seconds) to produce 1,000 MCMC draws with $N=25$ ($N=75$). 
Additional details related to computation are presented in the supplementary material.}

\subsection{Ozone data}
\label{sec:ozone}

Problems in the environmental sciences provide a broad area of application for which the 
proposed modeling approach is particularly well-suited. For such problems, 
it is of interest to estimate relationships between different
environmental variables, some of which are typically recorded on an
ordinal scale even though they are, in fact, continuous variables.
This is also a setting where it is natural to model the joint
stochastic mechanism for all variables under study from which 
different types of conditional relationships can be explored.

To illustrate the utility of our methodology in this context, we work
with data set {\tt ozone} from the ``ElemStatLearn'' R package.  
This example contains four variables: ozone concentration (ppb), 
wind speed (mph), radiation (langleys), and temperature (degrees 
Fahrenheit). While these environmental characteristics are all random 
and interrelated, we focus on estimating ozone concentration as a 
function of radiation, wind speed, and temperature. To apply our
model, rather than using directly the observed ozone concentration, 
we define an ordinal response containing three ozone concentration categories. 
Ozone concentration greater than 100 ppb is defined as high (ordinal level
$3$); this can be considered an extreme level of ozone concentration,
as only about $6\%$ of the total of 111 observations are this high. Concentration
falling between 50 ppb and 100 ppb (approximately $25\%$ of the 
observations) is considered medium (ordinal level $2$), and 
values less than $50$ ppb are assumed low (ordinal level $1$).  
We use this to illustrate a practical setting in which an ordinal
response may arise as a discretized version of a continuous
response. Moreover, this example enables comparison of inferences 
from the ordinal regression model with a model based on the actual continuous responses.

%
%In this example continuous ozone concentration is contained in the
%data, and we create a discretized response indicating ozone
%concentration level to illustrate the proposed method. However, in
%other real settings where data on ozone is recorded, it may only be
%available in an ordinal form, such as whether or not it exceeded a
%certain threshold, or whether it was high, medium, or low.  This idea
%can be generalized to other environmental characteristics or outcomes, 
%which may only be available on an ordinal scale, but in reality are continuous.
%

\begin{figure}[t]
\centering
\includegraphics[height=4.3in,width=4.9in]{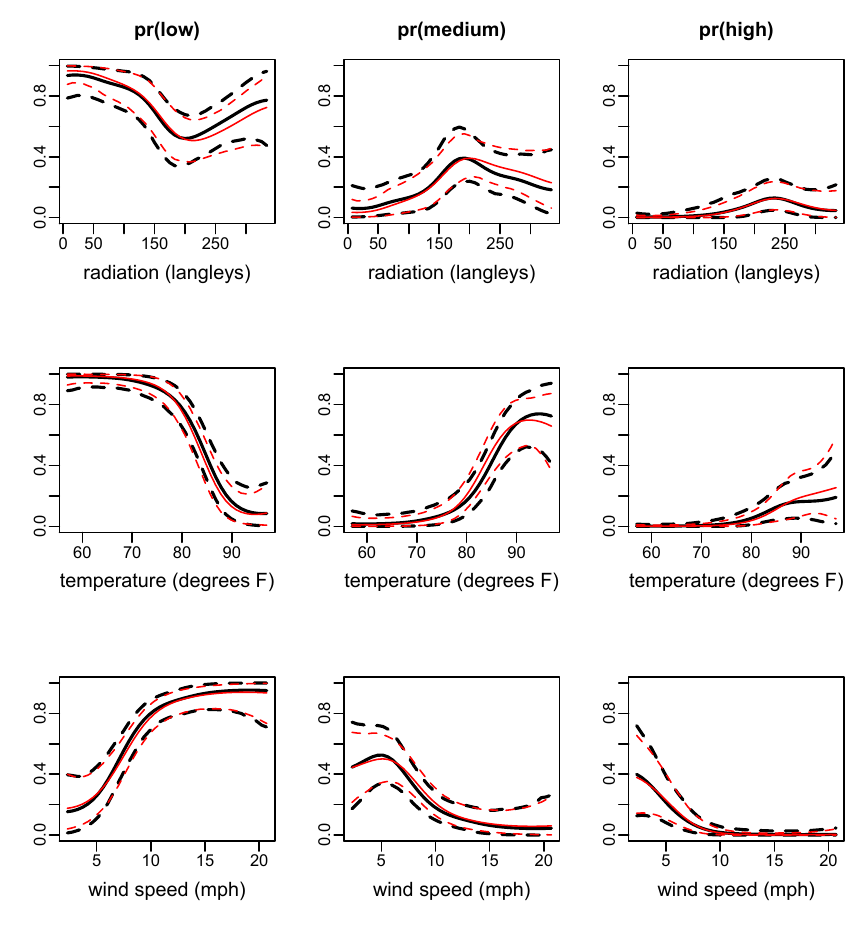}
\caption{Ozone data. Posterior mean (solid lines) and $95\%$ interval estimates
(dashed lines) for $\mathrm{Pr}(Y=l \mid  x_m,G)$ (thick black) compared to
$\mathrm{Pr}(\gamma_{l-1}<Z\leq \gamma_l \mid  x_m,G)$ (red), for
$l=1,2,3$ and $m=1,2,3$, giving the probability that ozone concentration 
is low, medium, and high over covariates radiation, temperature, and 
wind speed.}
\label{fig:ord_ozone}
\end{figure}

The model was applied to the ozone data, with response, $Y$, given by 
discretized ozone concentration, and covariates, $\boldsymbol{X}=$ 
$(X_{1},X_{2},X_{3})=$ (radiation, temperature, wind speed). 
To validate the inferences obtained from the DP mixture ordinal 
regression model, we compare results with the ones from a DP mixture 
of multivariate normals for the continuous vector $(Z,\boldsymbol{X})$, since 
the observations for ozone concentration, $Z$, are available on a
continuous scale.
The latter model corresponds to the density regression approach of
\cite{muller}, extended with respect to the resulting inferences in
\citet{taddy}. Here, it serves as a benchmark for our ordinal
regression model, since it provides the best possible inference that can
be obtained under the mixture modeling framework if no loss in
information occurs by observing $Y$ rather than $Z$.
We compare the univariate regression curves
$\mathrm{Pr}(Y=l \mid  x_m,G)$ with $\mathrm{Pr}(\gamma_{l-1}<Z\leq
\gamma_l \mid  x_m,G)$, for $l=1,2,3$, and $m=1,2,3$, the latter based
on the mixture model for $(Z,\boldsymbol{X})$. Figure \ref{fig:ord_ozone} 
compares posterior mean and $95\%$ interval estimates for the
regression curves given each of the three covariates. The key result
is that both sets of inferences uncover the same regression relationship 
trends. The only subtle differences are in the uncertainty bands which
are overall slightly wider under the ordinal regression model. Also 
noteworthy is the fact that the ordinal regression mixture model 
estimates both relatively standard monotonic regression functions 
(e.g., for temperature) as well as non-linear effects for radiation. 

\begin{figure}[t]
\begin{tabular}{ccc}
\includegraphics[height=1.9in,width=1.9in]{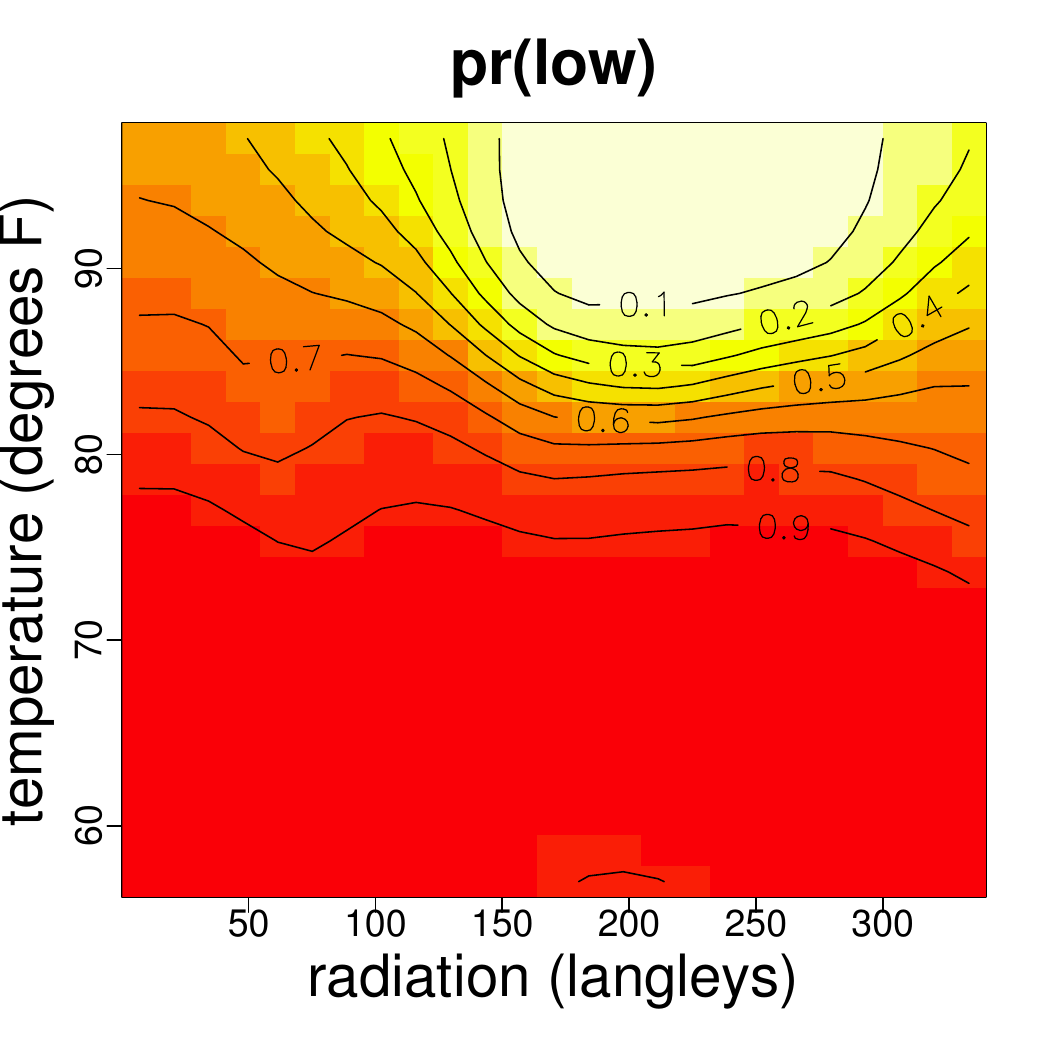}&
\includegraphics[height=1.9in,width=1.9in]{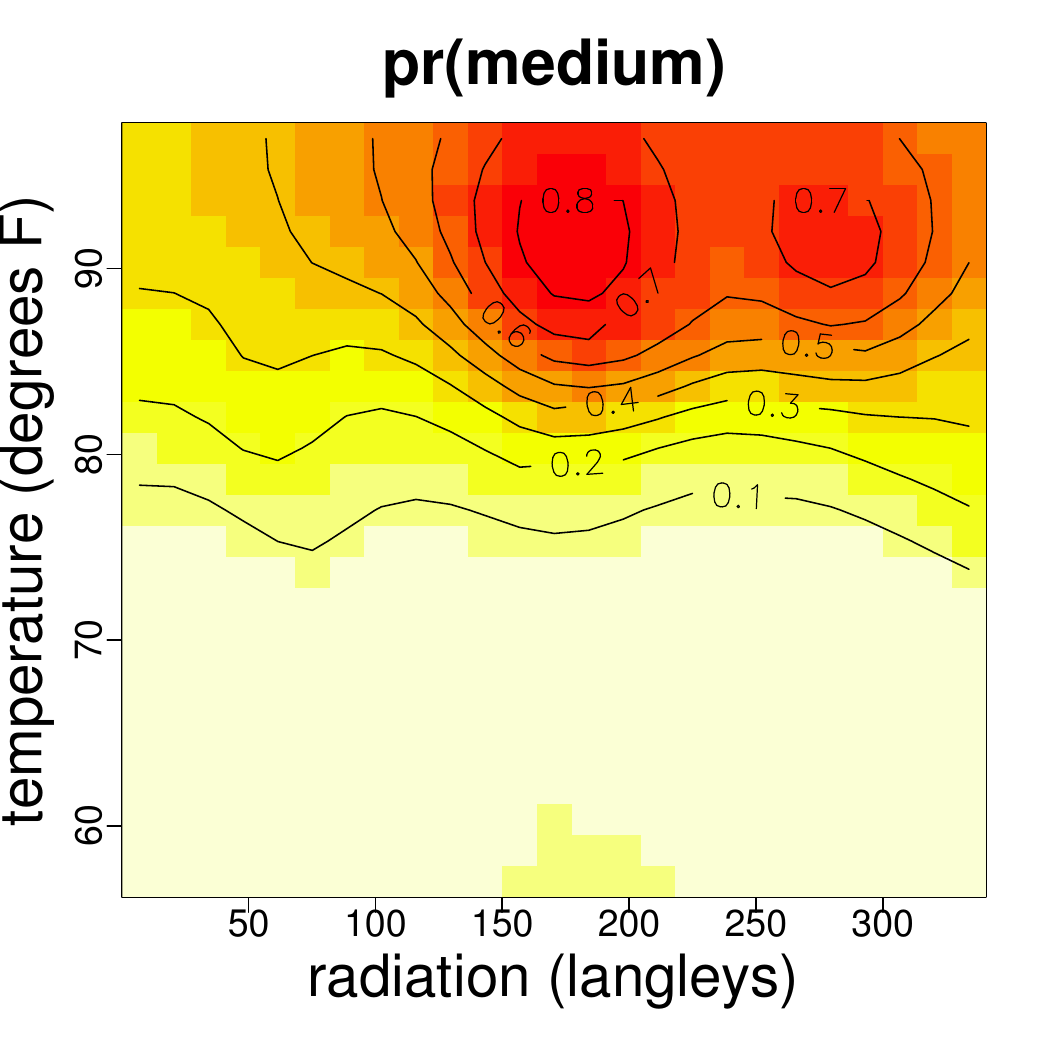}&
\includegraphics[height=1.9in,width=1.9in]{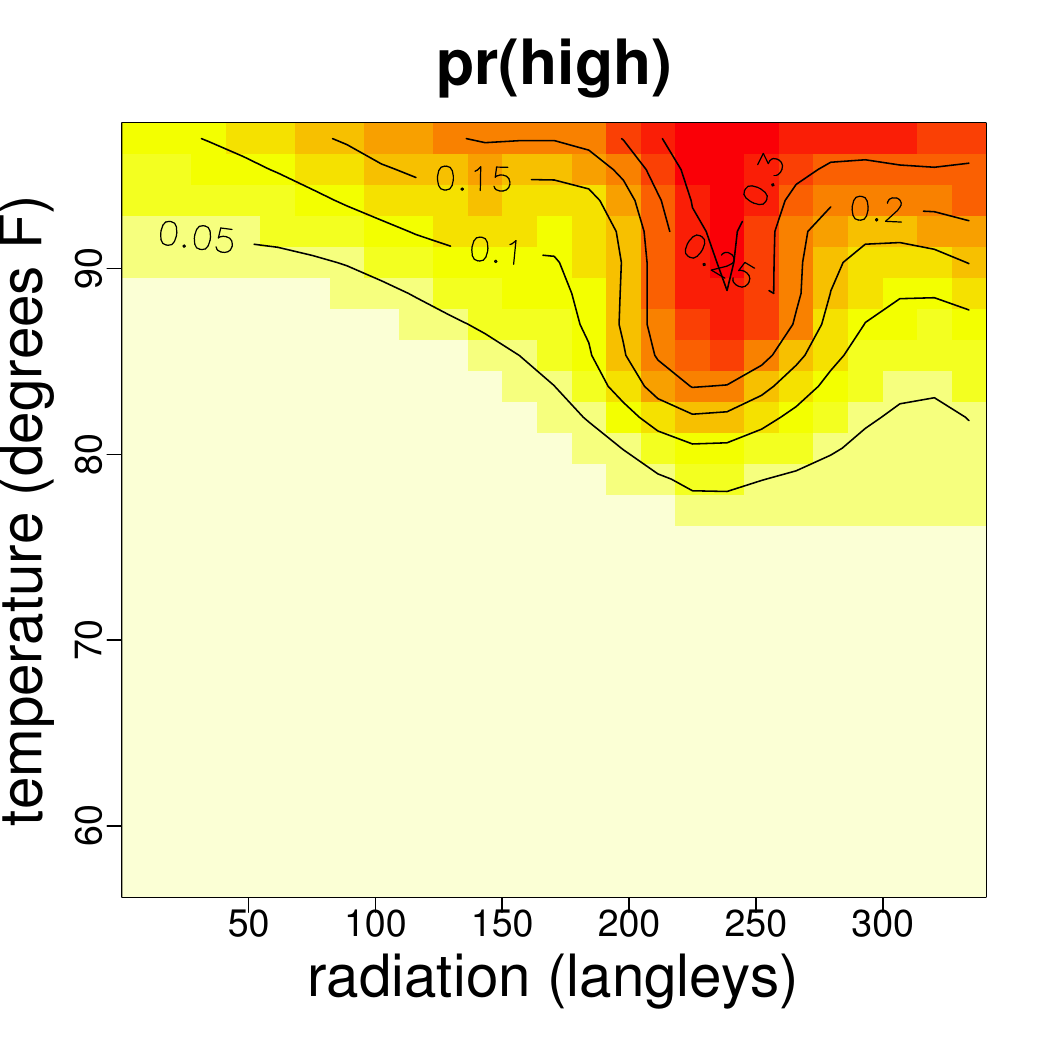}
\end{tabular}
\caption{Ozone data. Posterior mean estimates for $\mathrm{Pr}(Y=l \mid x_1,x_2,G)$, 
for $l=1,2,3$, corresponding to, from left to right,  low, medium, and
high ozone concentration. The spectrum of colors from white to red
indicates probabilities that range from 0 to 1.}
\label{fig:ord_ozone_biv}
\end{figure}

The ability to capture such a wide range of trends for the regression
relationships is a feature of the nonparametric mixture model. Another
feature is its capacity to accommodate interaction effects among the
covariates without the need to incorporate additional terms in the model.
Such effects are suggested by the estimates for the response
probability surfaces over pairs of covariates; for instance, Figure 
\ref{fig:ord_ozone_biv} displays these estimates as a function of 
radiation and temperature.

\subsection{Credit ratings of U.S. companies}
\label{sec:credit}

Here, we consider an example involving Standard and Poor's (S\&P) credit
ratings for $921$ U.S. firms in 2005. The example is taken from
\cite{verbeek}, in which an ordered logit model was applied to the
data, and was also used by \cite{chibgreen} to illustrate an additive
cubic spline regression model with a normal DP mixture error distribution. 
For each firm, a credit rating on a seven-point ordinal scale is available, 
along with five characteristics. Consistent with the analysis of \cite{chibgreen},
we combined the first two categories as well as the last two
categories to produce an ordinal response with $5$ levels, where
higher ratings indicate more creditworthiness. The covariates in this
application are book leverage $X_1$ (ratio of debt to assets),
earnings before interest and taxes divided by total assets $X_2$,
standardized log-sales $X_3$ (proxy for firm size), retained earnings
divided by total assets $X_4$ (proxy for historical profitability), 
and working capital divided by total assets $X_5$ (proxy for short-term liquidity).

\begin{figure}[t]
\centering
\includegraphics[height=3.5in,width=5.7in]{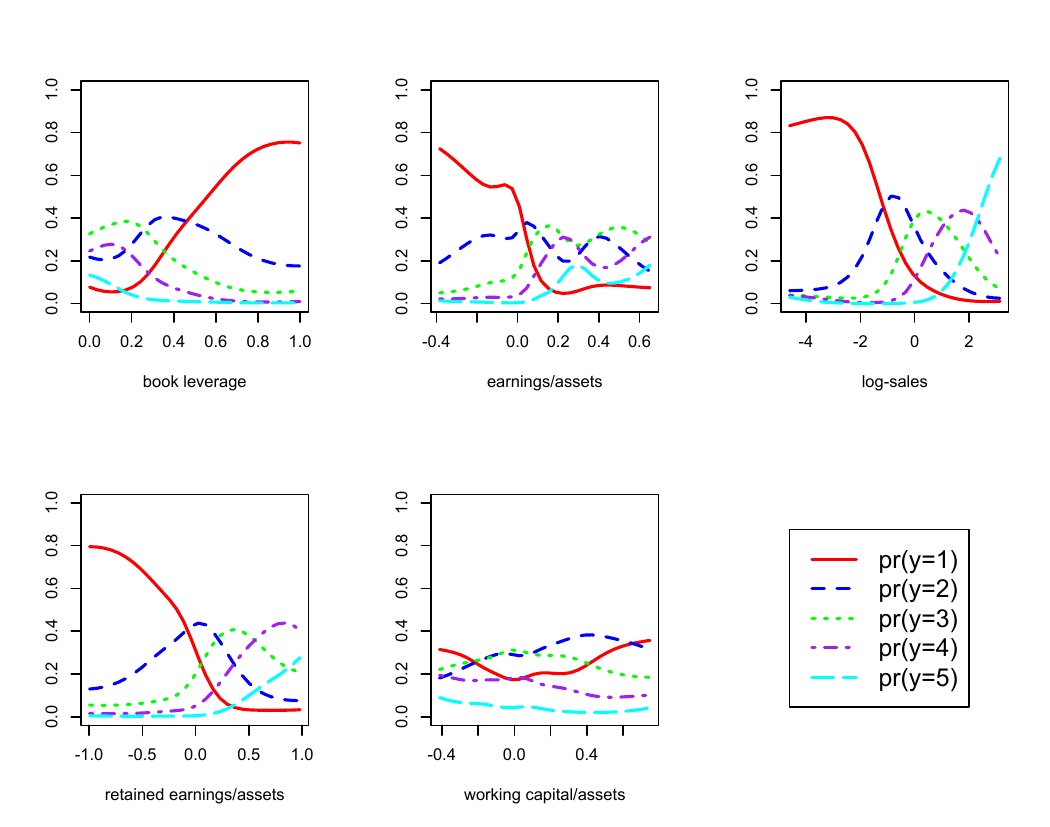}
\caption{Credit rating data. Posterior mean estimates for
$\mathrm{Pr}(Y=l \mid  x_m,G)$, for each covariate $m=1,\dots,5$. All
five ordinal response curves are displayed in a single panel for each
covariate.}
\label{fig:ord_credit_univ}\end{figure}

The posterior mean estimates for the marginal probability curves,
$\mathrm{Pr}(Y=l \mid  x_m,G)$, for $l=1,\dots,5$ and $m=1,\dots,5$,
are shown in Figure \ref{fig:ord_credit_univ}. 
%Each panel displays the set of regression functions associated with a single covariate.
These estimates depict some differences from the corresponding ones
reported in \cite{chibgreen}, which could be due to the additivity assumption
of the covariate effects in the regression function under their model. 
Empirical regression functions -- computed by calculating proportions 
of observations assigned to each class over a grid in each covariate -- 
give convincing graphical evidence that the regression relationships estimated 
by our model fit the data quite well.   

%
%The most nonstandard trends appear to be present over $X_2$, which is
%earnings before interest and taxes divided by total assets. The
%covariate $X_3$ (log-sales) has some interesting as well as sensible
%probability trends associated with it.  
To discuss the estimated regression trends for one of the covariates,
consider the standardized log-sales variable, which is a proxy for
firm size. The probability of rating level $1$ is roughly constant 
for low log-sales values, and is then decreasing to $0$, indicating 
that small firms have a similar high probability of receiving the 
lowest rating, whereas the larger the firm, the closer to $0$ this 
probability becomes. The probability curves for levels $2$, $3$,
and $4$ are all quadratic shaped, with peaks occurring at larger 
log-sales values for higher ratings. Finally, the probability of 
receiving the highest rating is very small for low to moderate 
log-sales values, and is increasing for larger values.
In summary, log-sales are positively related to credit rating, as expected.

\begin{figure}
\includegraphics[height=2.5in,width=6in]{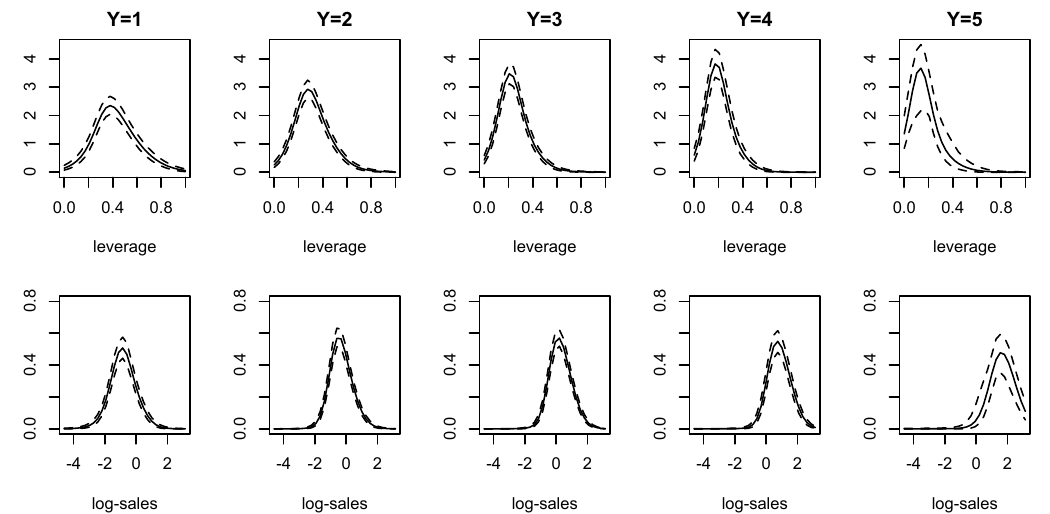}
\caption{Credit rating data. Posterior mean (solid lines) and $95\%$ interval 
estimates (dashed lines) for covariate densities $f(x \mid Y=l,G)$
conditional on ordinal credit rating $l=1,...,5$. The top row corresponds 
to covariate book leverage, and the bottom row to standardized log-sales.}
\label{fig:ord_credit_cov}
\end{figure}

This is another example where it is arguably natural to model
the joint distribution for the response and covariates (the specific
firm characteristics). This allows our model to accommodate
interactions between covariates, as we do not assume additivity 
or independence in the effects of the covariates on the latent 
response. In addition to the regression curve estimates, we may 
obtain inference for the covariate distribution, or for any covariate 
conditional on a specific ordinal rating. These inverse relationships
(discussed in Section \ref{sec:post}) could be practically relevant in
this application. It may be of interest to investors and
econometricians to know, for example, approximately how large is a
company's leverage, given that it has a rating of $2$? Is the
distribution of leverage much different from that of a level $3$
company? Figure \ref{fig:ord_credit_cov} plots the estimated 
densities for book leverage and standardized log-sales 
conditional on each of the five ordinal ratings. In general,
the distribution of book leverage is centered on smaller values as
rating increases, and the densities become more peaked supporting a
smaller range of leverage values for higher ratings. The interval
bands are slightly wider for the distribution associated with $Y=1$
than for $Y=2$, $3$, or $4$, and they are much wider for $Y=5$, 
which is consistent with the small number of firms with a rating of $5$.
The distribution of log-sales has a mode which occurs at increasing 
values as rating increases, indicating that if one firm has a higher 
rating than another, it likely also has higher sales.

\subsection{Standard and Poor's grades of countries}
\label{sec:countries}

As a second econometrics example, we consider a data
set from \cite{simonoff}, comprising S\&P ratings of $n=31$ 
countries along with debt service ratio and income, the latter recorded 
on an ordinal scale with levels of low, medium, and high. Ratings range 
from 1 to 7, with 1 indicating the best rating of AAA, and 7 the worst 
of CCC. With two covariates and a very small sample size, this example
provides an interesting testbed for our modeling approach.

Since income is available as a discrete variable, $W$, we model it 
through the (latent) continuous income variable, $Z_{2}$. 
%\citep[this method of modeling ordinal covariates was also used by][]{ronning}. 
Therefore, $X$ represents debt service ratio, and $\boldsymbol{Z}=$ 
$(Z_1,Z_2)$, where $W$ arises from $Z_2$ just as the ordinal rating 
response $Y$ arises through latent continuous response $Z_1$. 
This is another application where one may be interested in inverse 
relationships, such as the distribution of debt service ratio and/or 
income given a specific S\&P rating.

\begin{figure}
\begin{tabular}{cccc}
\includegraphics[height=1.4in,width=1.4in]{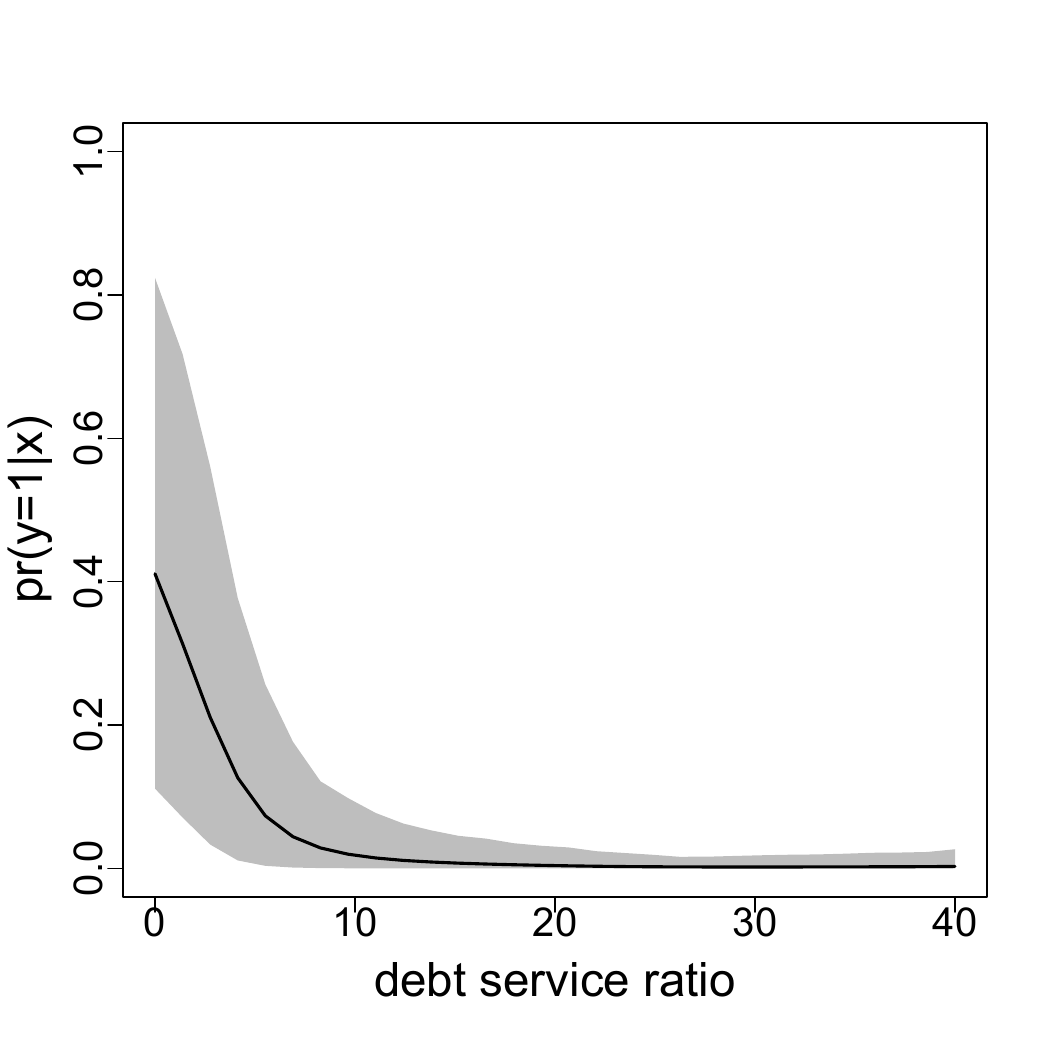}&
\includegraphics[height=1.4in,width=1.4in]{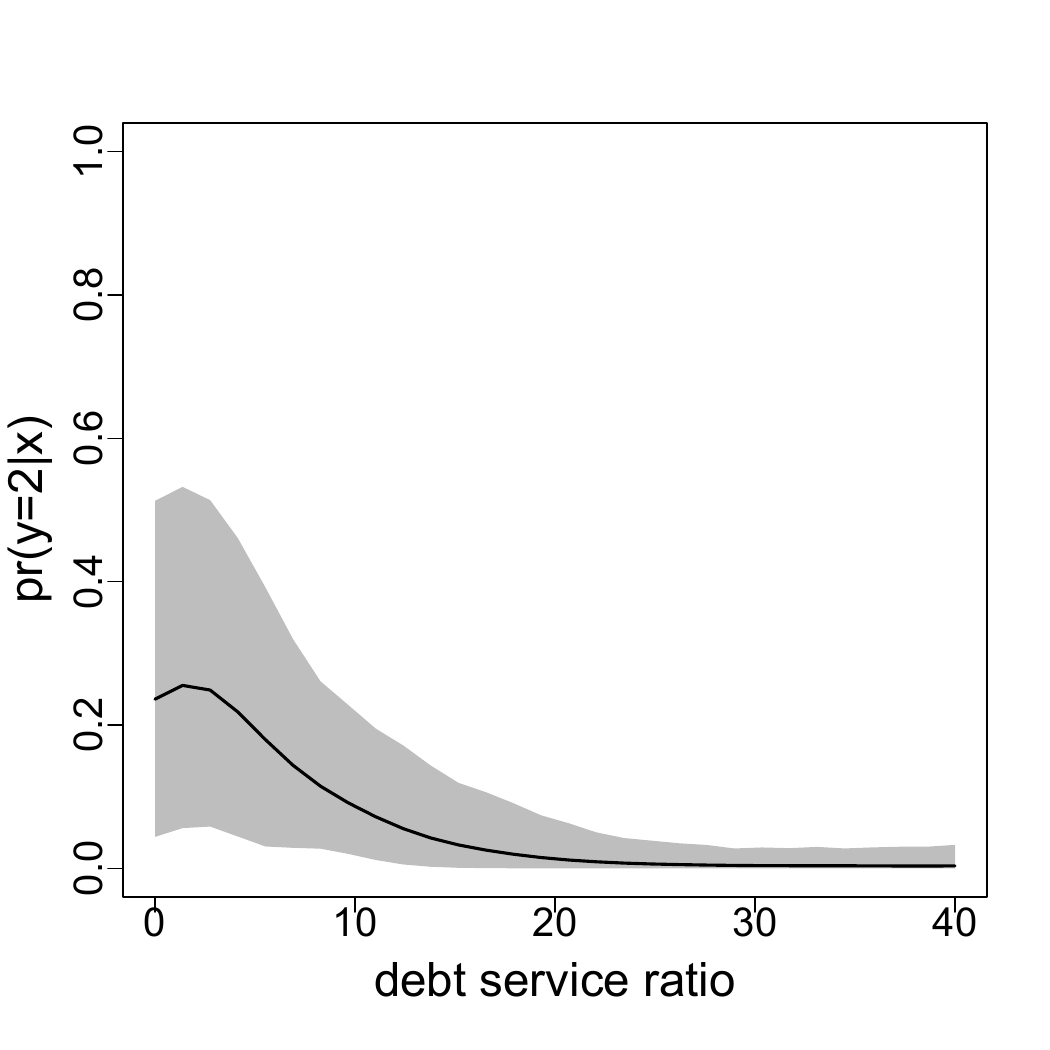}&
\includegraphics[height=1.4in,width=1.4in]{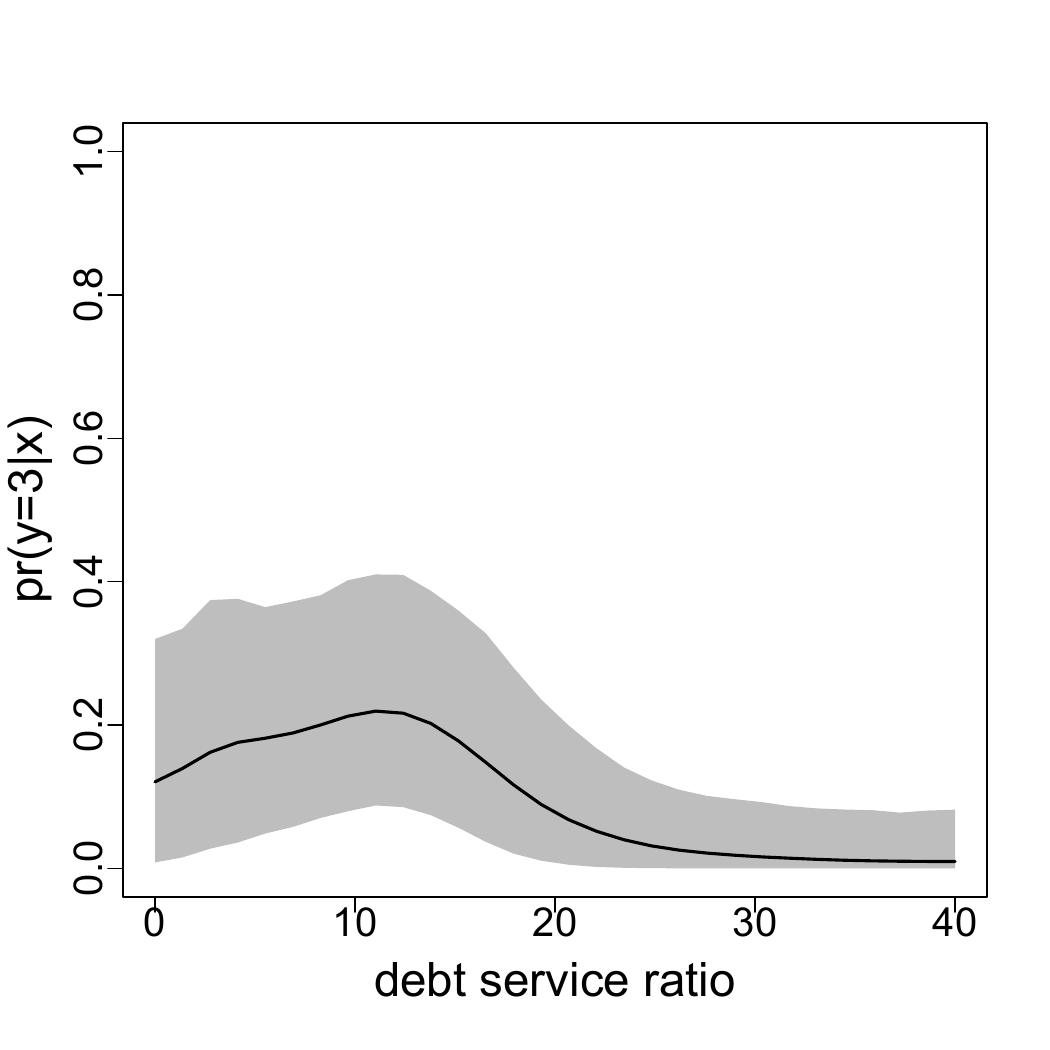}&
\includegraphics[height=1.4in,width=1.4in]{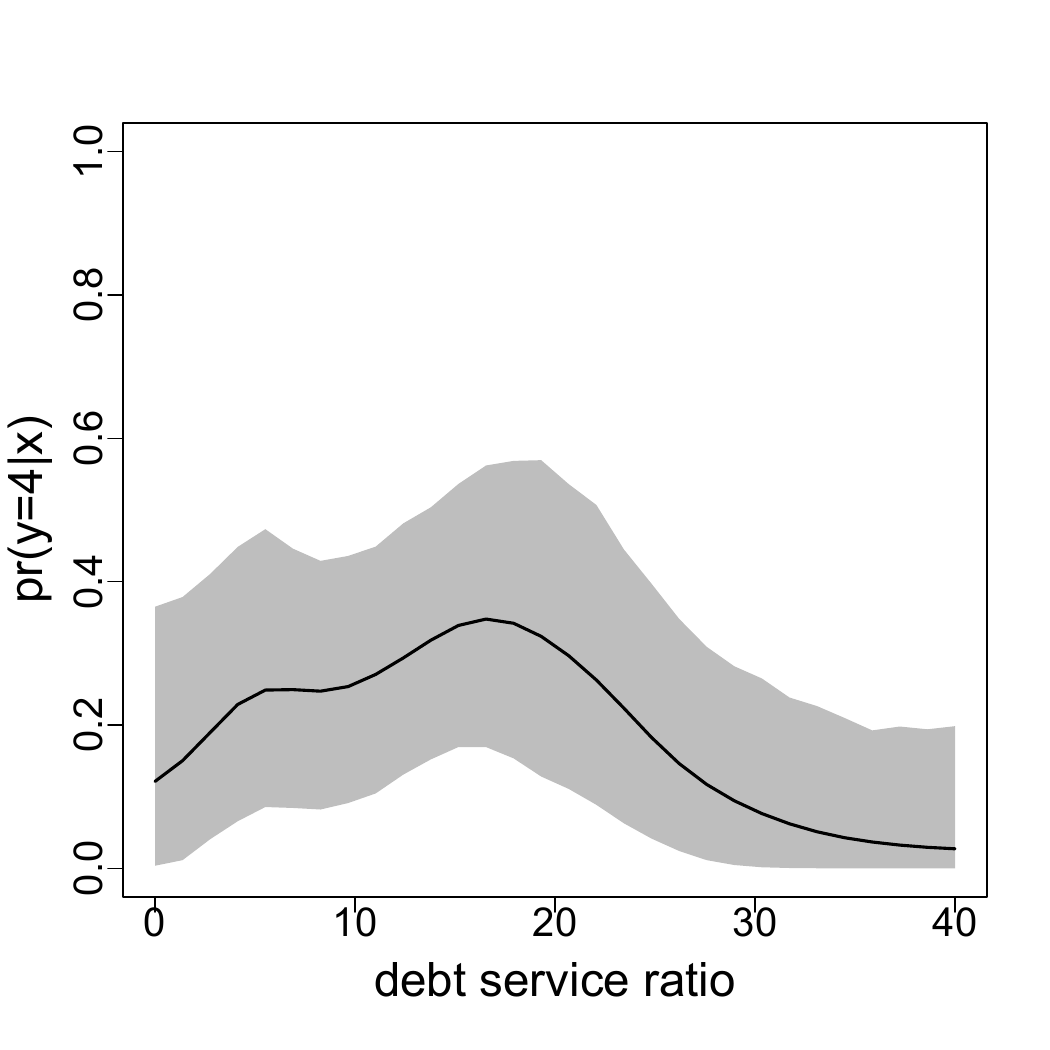}
\end{tabular}

\centering

\begin{tabular}{ccc}
\centering
\includegraphics[height=1.4in,width=1.4in]{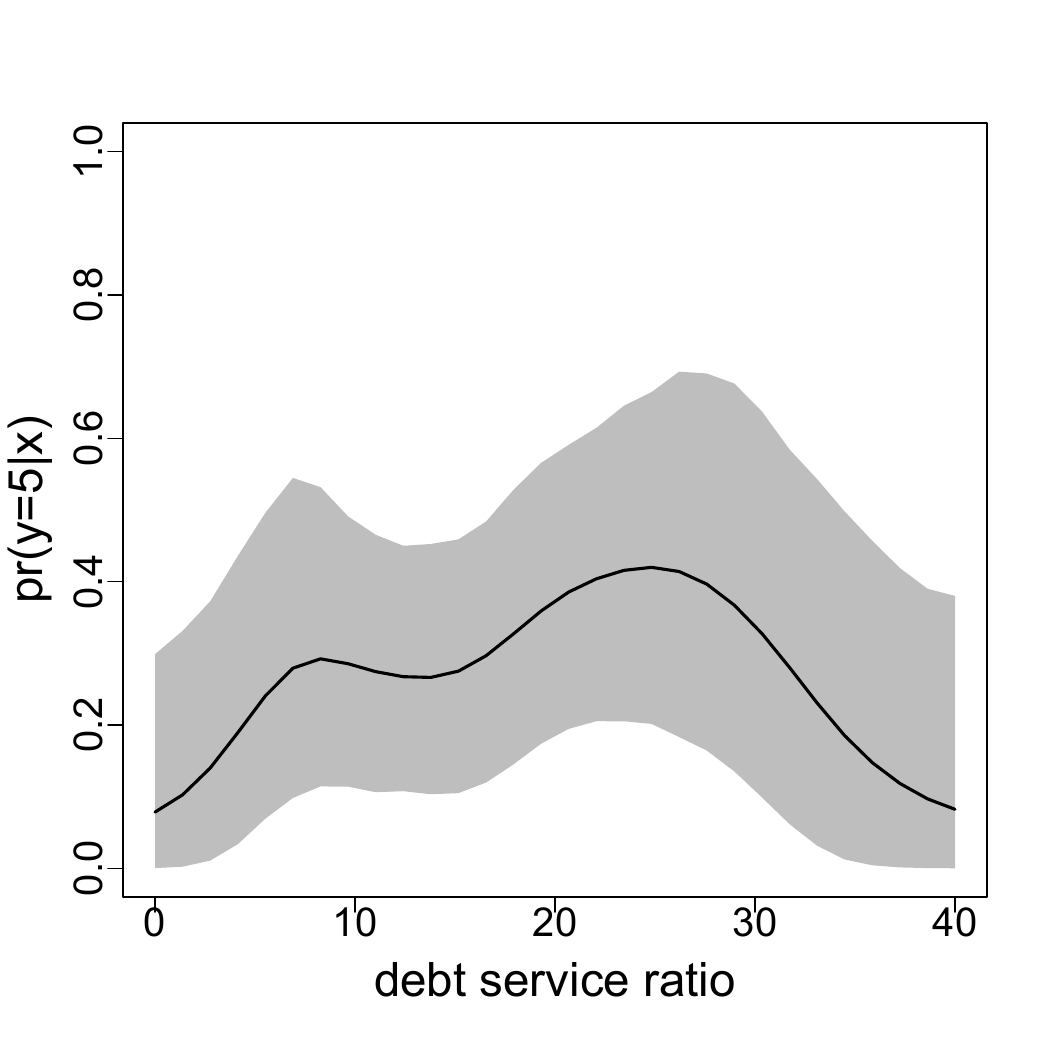}&
\includegraphics[height=1.4in,width=1.4in]{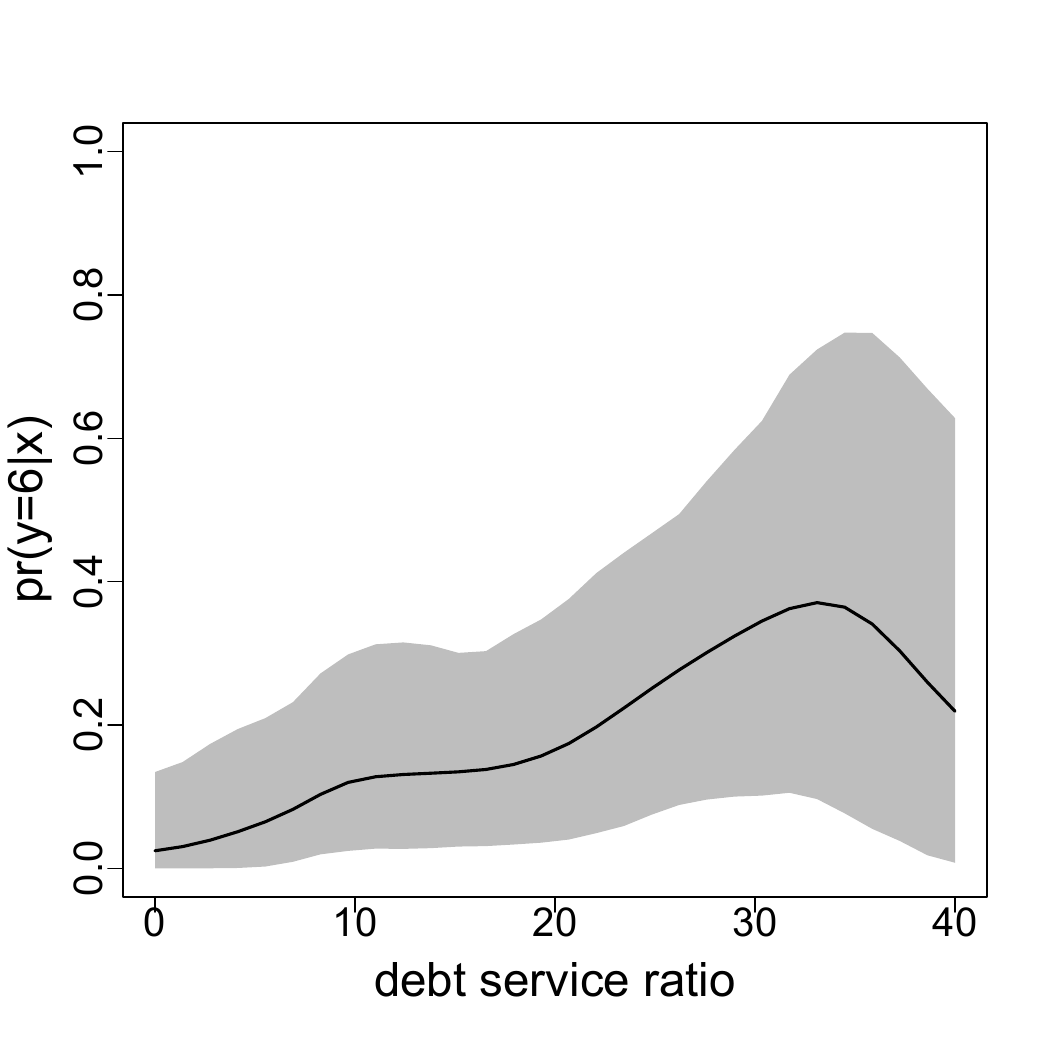}&
\includegraphics[height=1.4in,width=1.4in]{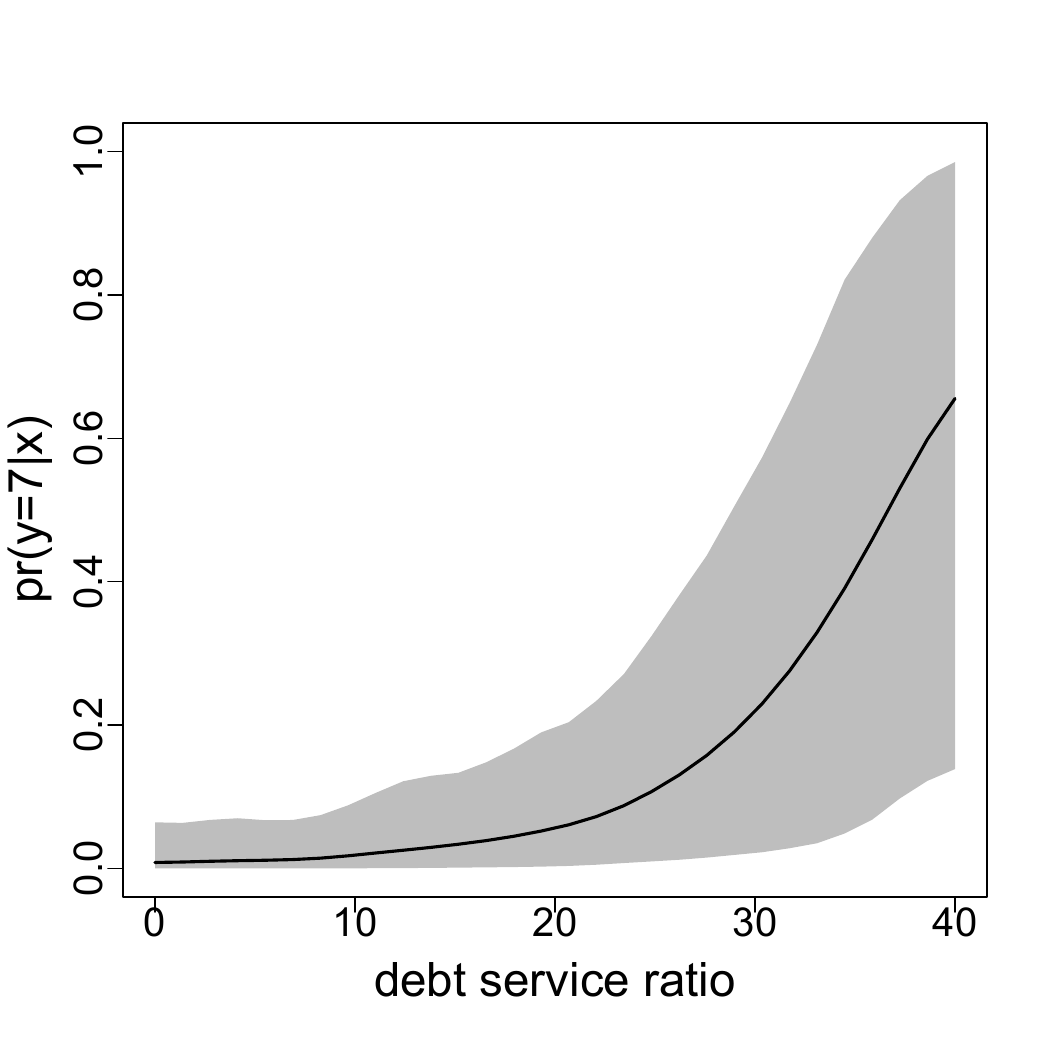}
\end{tabular}
\caption{S\&P ratings of countries data. Posterior mean (solid lines) and $95\%$ intervals 
(gray bands) for the probability response curve associated with each
rating, ranging from AAA (level 1) to CCC (level 7), as a function of debt service ratio.}
\label{fig:sp_ord_ratings}
\end{figure}

The probability response curves as functions of debt service ratio
(Figure \ref{fig:sp_ord_ratings}) contain both monotonic trends
(decreasing for response categories 1 and 2, and increasing for 
category 7), as well as non-linear ones, most notably for categories 
4 and 5. The interval bands are wider than in earlier examples, 
given the smaller sample size. Although not shown here, regression 
curves can also be obtained over discrete income from 
$\mathrm{Pr}(Y=j \mid W=w,G)=$ 
$\mathrm{Pr}(Y=j,W=w \mid G)/\mathrm{Pr}(W=w \mid G)$, for $w=1,2,3$ (low,
medium, and high income), where the numerator contains a double integral of a
bivariate normal density function. The probability of receiving a top 
rating of 1, 2, or 3 is highest for high-income countries, the
probability of receiving a moderate rating of 4 or 5 is highest for 
medium-income countries, and the probability of receiving a poor 
rating is highest for low-income countries. It is highly unlikely for 
a country to receive one of the top two ratings unless it is high-income, 
however there is non-negligible probability of a medium-income country 
receiving one of the two lowest ratings.

\setlength{\tabcolsep}{1pt}

\begin{figure}[t!]
\centering
\includegraphics[height=2.3in,width=2.5in]{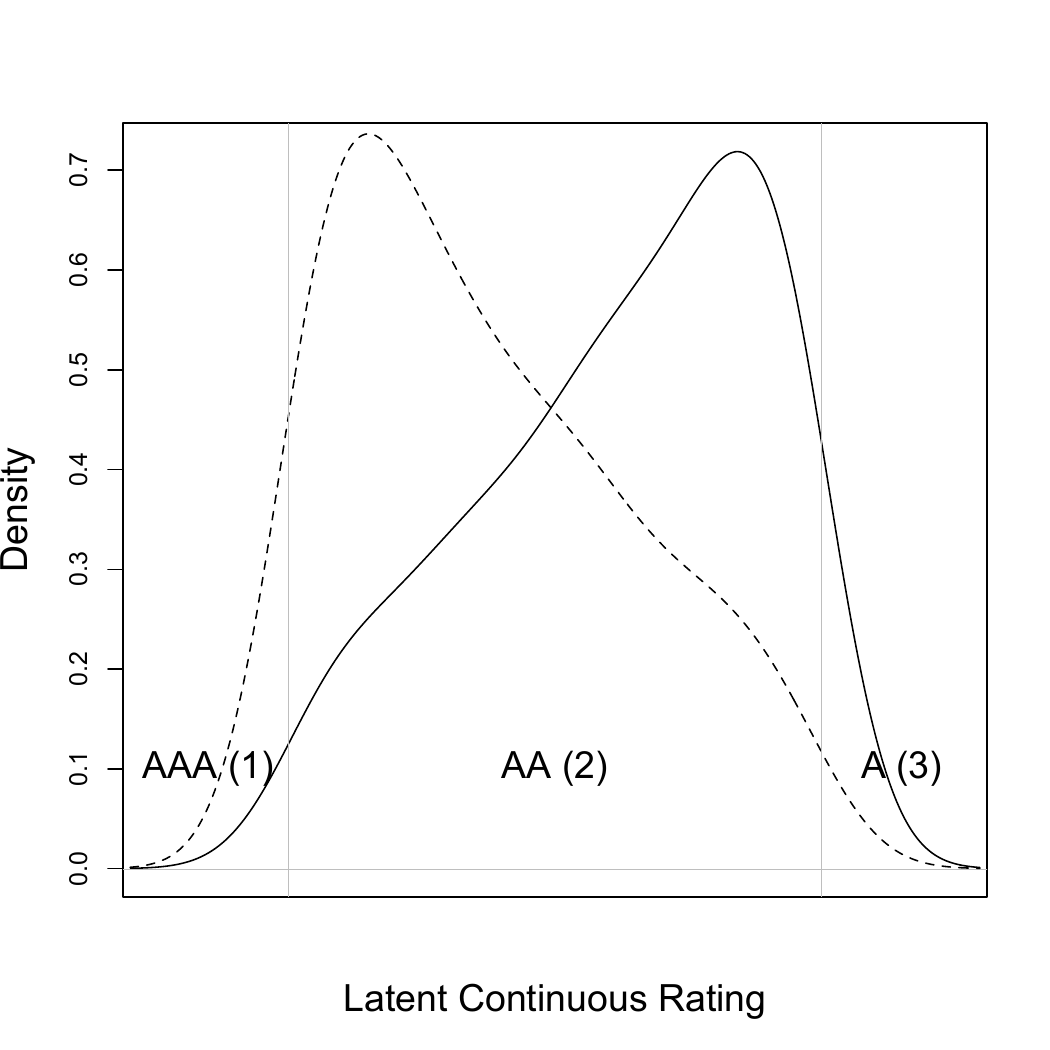}
\includegraphics[height=2.3in,width=2.5in]{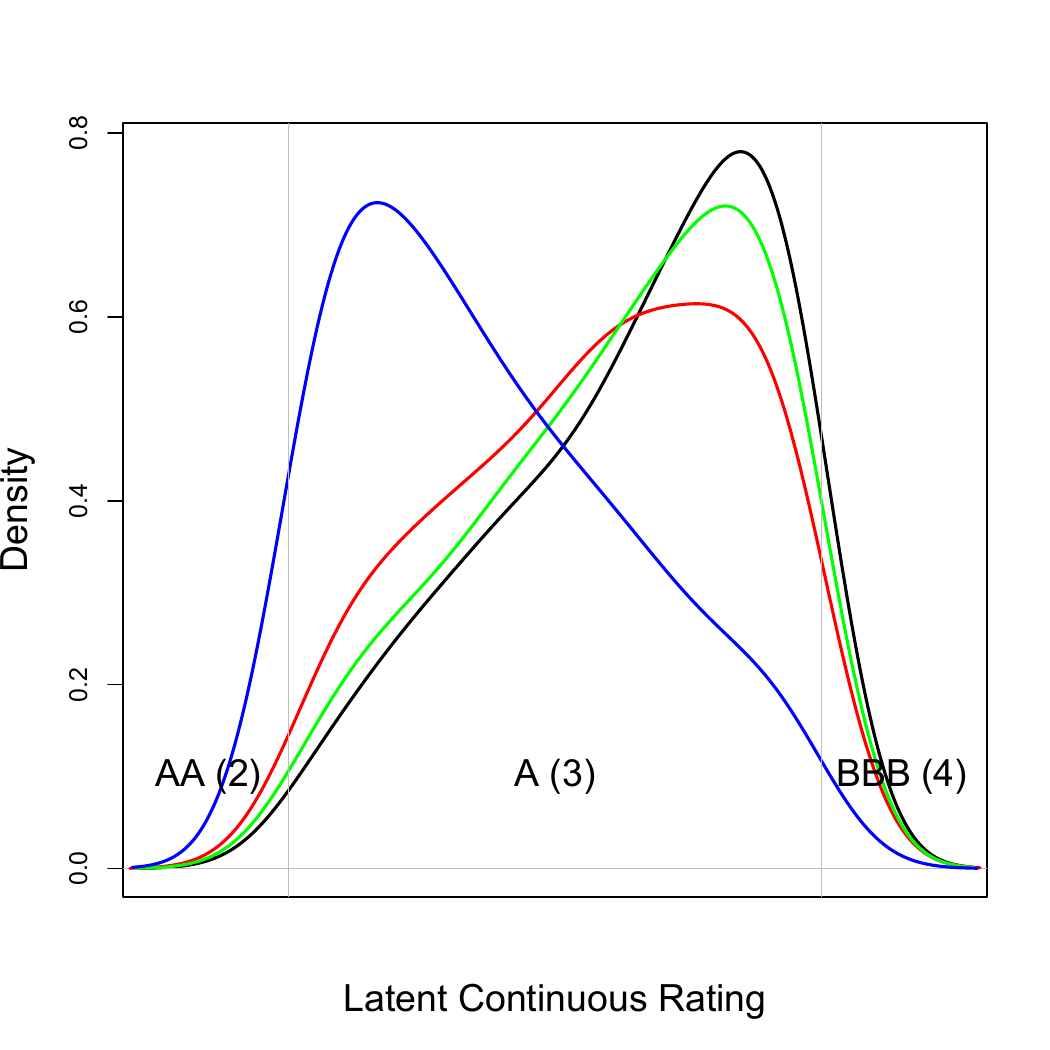}
\caption{S\&P ratings of countries data. Left panel: Posterior densities
of latent continuous ratings for Australia (solid line) and Canada
(dashed line), the two countries with AA rating. Right panel:
Posterior densities of latent continuous ratings for the four
countries with A rating: Chile (black), Czech Republic (red), 
Hungary (green), and Slovenia (blue). In both panels, the gray 
vertical lines indicate the borders for ordinal ratings.}
\label{fig:ord_sp_latent}
\end{figure}

The latent continuous responses represent latent continuous credit
rating in this application.  The method for posterior simulation
involves sampling $z_{i,1}$, for $i=1,\dots,31$, which represent the
country-specific latent ratings. The two countries with AA (level 2)
rating are Canada and Australia. Both of these countries have income
classified as high, however Canada has no debt, whereas Australia has a
debt service ratio which is around 10. This value is not particularly
high, but since higher debt service ratio seems to be associated with
poorer ratings, we would expect that Canada would be closer to
receiving a better rating of AAA than Australia. Indeed, posterior
densities of latent continuous ratings for Canada and Australia
(Figure \ref{fig:ord_sp_latent}, left panel) indicate that Canada's ordinal 
rating of AA is closer to a AAA rating than Australia's.

Next, consider the four countries with A rating (level 3): Chile, Czech
Republic, Hungary, and Slovenia. The first three of these countries are
classified as medium income, whereas Slovenia is a high income
country. The debt service ratios range from 8.9 (Czech Republic) to 15.8 (Chile).
The estimated latent response densities are shown on the right panel 
of Figure \ref{fig:ord_sp_latent}.  We note that Slovenia's latent rating
distribution is centered on values close to the cut-off point for a
better rating of AA. The other three distributions are fairly similar. 
Chile appears closest to a BBB rating, which is consistent 
with its higher debt service ratio. An interesting observation from 
these results is that differences in income appear to have a larger
effect on the latent rating distributions than differences in debt service ratio.

\subsection{Analysis of multirater agreement data}
\label{sec:multirater}

A variety of methods exist for analyzing ordinal data collected from multiple raters when the goal is to measure 
agreement. Such methods 
range from the commonly used $\kappa$ statistic \citep{cohen} and its extensions \citep{fleiss}, 
which are indices calculated from the observed and expected agreement of raters, to model-based approaches 
involving log-linear models \citep{tanner}.  We do not attempt a comprehensive 
review here, rather our focus lies in the use of model-based methods for ordinal responses 
collected from multiple raters along with covariate information. The proposed multivariate ordinal regression 
model offers flexibility in this setting in terms of the modeling framework and resulting inferences.  
We focus on a scenario involving a set of expert graders who evaluate student essays, rating them on an ordinal 
scale. We contrast our approach to the parametric model of
\cite{johnson}, from where the specific data example is taken, 
and the semiparametric approach of \cite{dalal}, both developing Bayesian inference built from modeling for 
latent responses.

Multirater agreement data arises when $k$ raters assign ordinal scores to $n$ individuals, such 
that $\boldsymbol{y}_i=$ $(y_{i1},\dots,y_{ik})$ collects all scores for the $i$th individual. 
The raters typically use the same classification levels, and therefore each $y_{ij}\in \{1,\dots,C\}$.  
This data structure could be summarized in a contingency table, however, we are concerned with 
problems in which relevant covariate information is available for each individual.  
We assume that each judge assigns an ordinal rating to individual $i$, which represents a discretized 
version of a continuous rating; that is, $y_{ij}$ is determined by $z_{ij}$, the continuous latent score 
assigned by judge $j$ on individual $i$.

This is in contrast to the formulation of \cite{johnson}, where all
judges are assumed to agree on the intrinsic worth of each item, 
such that $z_{ij}=w_i+\epsilon_{ij}$, where $w_i$ represents the true 
latent score, and $\epsilon_{ij}$ is the error observed by judge $j$. 
Then, $w_i$ is linearly related to the covariates, assumed normal 
with mean containing the term
$\boldsymbol{x}_i^T \boldsymbol{\beta}$. The normal latent 
response distribution is not appropriate when grade distributions 
are skewed or favor low/high scores over moderate scores. 
Since the distribution of the $z_{ij}$ does not have a judge-specific 
mean, random cut-offs are necessary to allow the ratings of a 
particular subject to vary among judges.

\cite{dalal} note that the assumption of intrinsic agreement among 
raters may be inappropriate when raters have different beliefs or 
perspectives that may influence their scoring behavior. They assume a 
DP mixture model for the judge-specific latent random vectors built
from a mixture kernel defined through independent normals. Dependence 
is therefore introduced over the latent scores of a single rater
(albeit under a restrictive product-kernel for the mixture), but the data
vectors arising from each rater are assumed independent. Under this 
model, it is therefore unclear how to extract inference for inter-rater 
agreement, which is a key inferential objective for applications of
this type in the social sciences.

Similar to the approach in \cite{dalal}, there is no notion in our
model of an intrinsic true score for an individual.
%since we assume each rater has his or her own beliefs which determine
%the score assigned to a particular individual. 
An overall score for an individual could be obtained by averaging in 
some fashion over the latent scores assigned by each rater.
%however, extracting a true underlying score which it is believed all
%raters agree on is not the main goal here. Rather, we focus on making 
However, our main goal here is to obtain inference about
relationships between the ordinal scores and the covariates, 
as well as for inter-rater agreement over both the covariate space 
and the scores. Our method offers a potentially useful perspective 
for modeling multirater agreement data, most notably with respect to 
the generality of the nonparametric mixture model which can 
accommodate complex dependence among raters and non-linear 
relationships with the random covariates.

We apply our method to a problem involving three expert graders
who evaluate $n=198$ student essays, each assigned a rating on an
ordinal scale of $1$ through $10$ (these represent raters 2, 3, and 4
from the data given in Chapter 5 of \citealp{johnson}). 
Average word length and total number of essay words 
are used as the $p=2$ covariates, to study if they have an effect on
grader ratings.
%which may or may not explain to some extent the ratings given 
%by a particular judge. 
The traditional measure of agreement between raters $l$ and $m$ in 
the social sciences, the polychoric correlation $\rho_{lm}=$ $\text{corr}(Z_l,Z_m)$,
can be assessed through the covariance mixing parameters
$(\boldsymbol{\Sigma}_1,\dots,\boldsymbol{\Sigma}_N)$.
As discussed in Section \ref{sec:post}, the posterior predictive
distribution for $\rho_{lm}$ can  be obtained by sampling at each MCMC
iteration the corresponding $(\rho_{lm,1},\dots,\rho_{lm,N})$ 
with probabilities $(p_1,\dots,p_N)$. 
%These distributions suggest strongest agreement between raters 1 and
%3, and similar levels of agreement between the other two pairs of
%raters. All three distributions place some probability on negative
%correlations, suggesting there is some disagreement present between
%all pairs of raters.  
The polychoric correlation predictive distributions for all three pairs
of raters favor more heavily positive correlations (raters 1 and 3
appear to agree most strongly), but place substantial probability 
on negative correlations.  
We can study where raters $l$ and $m$ tend to agree or disagree by
grouping the latent continuous ratings according to the strength and
direction of $\text{corr}(Z_l,Z_m)$. For instance, a plot of 
$\text{E}(z_{il} \mid \text{data})$ and $\text{E}(z_{im} \mid \text{data})$ 
arranged by $\text{E}(\text{corr}(z_{il},z_{im})\mid\text{data})$
%as $\{\boldsymbol{\Sigma_l}: l=1,\dots,N\}$, and $L_i$ imply a particular
%$\text{corr}(z_{i,l},z_{i,m})$. 
reveals that raters 1 and 2 strongly agree on very low
ratings, but disagree when rater 2 assigns low ratings and rater 1
high ratings. It is also the case for the other pairs of raters that
they strongly agree mainly at low scores.

\begin{figure}[t!]
\centering
\includegraphics[height=2.3in,width=6in]{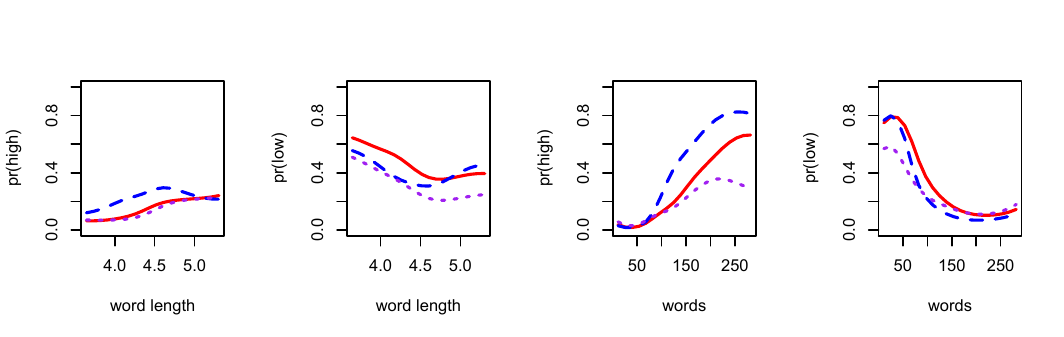}
\caption{Multirater data. Posterior mean estimates for probability of
high and low rating as a function of average word length (two left
plots) and total number of words (two right plots), for raters 1 (solid
red lines), 2 (dashed blue lines), and 3 (dotted purple lines).}
\label{fig:ord_multi_high_low}
\end{figure}

The model provides a variety of inferences in this multivariate ordinal 
regression example, which can be used to assess how ratings vary across
covariates, as well as how raters behave in comparison to one another.
Defining a high rating as 8 or higher, and a low rating as 3 or lower,
Figure \ref{fig:ord_multi_high_low} plots estimates for the
probability of high and low rating in terms of average word 
length and total number of words. There appears to be a strong
trend in rating as a function of number of words for each rater, with
rater 2 in particular assigning higher ratings for essays with more
words. The regression curves for high ratings associated with rater 2
are somewhat separated from raters 1 and 3, suggesting that 
overall rater 2 assigns higher ratings.

\begin{figure}[t!]
\centering
\includegraphics[height=2in,width=5.4in]{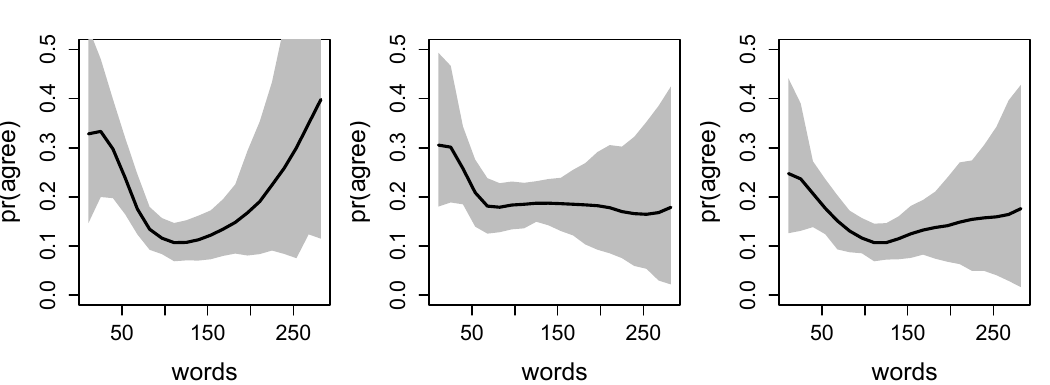}
\caption{Multirater data. Posterior mean (solid lines) and $95\%$ interval 
estimates (gray bands) for probability of agreement as a function of
total number of words. The left panel corresponds to raters 1 and 2,
the middle panel to raters 1 and 3, and the right panel to raters 2 and 3.}
\label{fig:ord_multi_agree}
\end{figure}

To identify regions of the covariate space in which raters tend
to agree or disagree, Figure \ref{fig:ord_multi_agree} plots estimates
for the probability of perfect agreement for the three pairs of raters
as a function of total number of words. This inference suggests that 
raters 1 and 2 agree most strongly on grades for essays with few or 
many words. 
%The other two pairs of raters tend to agree most for essays with few words, and
The trend in probability of agreement is weaker for the other two 
pairs of raters.

Finally, to assess the strength of agreement between raters on high
and low scores, we study the probability that one rater gives a 
high/low rating, conditional on the rating given by another rater. 
Table \ref{tab:ord_multi_agree_table} includes posterior means 
for $\mathrm{Pr}(Y_l \mid Y_{m},G)$, for $l,m \in \{1,2,3\}$, with 
$Y_l$ and $Y_m$ taking values of $\{8,9,10\}$ (high) or $\{1,2,3\}$ (low). 
Each row represents the event being conditioned on, while each column 
represents the event a probability is being assigned to. For example, 
row 1, column 3 contains the posterior mean for 
$\mathrm{Pr}(Y_2\in \{8,9,10\}\mid  Y_1\in \{8,9,10\},G)$. The cells
corresponding to disagreement are highlighted with gray. The first two
rows give probabilities conditional on rater 1 scores, and indicate that
rater 2 has more disagreement with rater 1 than does rater 3. The last 
two rows suggest that rater 2 disagrees more with rater 3 than does
rater 1. Finally, from the middle two rows, we note slightly more 
disagreement between raters 1 and 2 than between raters 3 and 2.

\setlength{\tabcolsep}{3pt}
\begin{table}[t!]
\centering
\begin{tabular}{ r | c | c | | c | c | | c |c |  }
\multicolumn{1}{r}{}
 &  \multicolumn{1}{c}{$Y_1=\mathrm{H}$}
 & \multicolumn{1}{c}{$Y_1=\mathrm{L}$}
 & \multicolumn{1}{c}{$Y_2=\mathrm{H}$}
 & \multicolumn{1}{c}{$Y_2=\mathrm{L}$}
 & \multicolumn{1}{c}{$Y_3=\mathrm{H}$}
 & \multicolumn{1}{c}{$Y_3=\mathrm{L}$} \\
\cline{2-7}
$Y_1=\mathrm{H}$ &  &  & $0.54$ &  \cellcolor{Gray}$0.15$ & $0.46$ &  \cellcolor{Gray}$0.06$ \\
\cline{2-7}
$Y_1=\mathrm{L}$ &  &  &  \cellcolor{Gray}$0.13$ & $0.48$ &  \cellcolor{Gray}$0.04$ & $0.51$ \\
\cline{2-7}
$Y_2=\mathrm{H}$ & $0.36$ &  \cellcolor{Gray}$0.18$ &  &  & $0.27$ &  \cellcolor{Gray}$0.14$ \\
\cline{2-7}
$Y_2=\mathrm{L}$ &  \cellcolor{Gray}$0.08$ & $0.56$ &  &  &  \cellcolor{Gray}$0.07$ & $0.41$ \\
\cline{2-7}
$Y_3=\mathrm{H}$ & $0.63$ &  \cellcolor{Gray}$0.11$ &  $0.55$ &  \cellcolor{Gray}$0.18$ &  &  \\
\cline{2-7}
$Y_3=\mathrm{L}$ & \cellcolor{Gray}$0.04$ & $0.78$ &  \cellcolor{Gray}$0.15$ & $0.54$ &  &  \\
\cline{2-7}
\end{tabular}
\caption{Multirater data. Posterior means for agreement and disagreement 
conditional probabilities for pairs of raters, with disagreement highlighted in gray 
(see Section \ref{sec:multirater} for details). H refers to high
ratings of $\{8,9,10\}$, and L refers to low ratings of $\{1,2,3\}$.}
\label{tab:ord_multi_agree_table}
\end{table}

%
%-----------------------------------------------------------------------
%

\section{Discussion}
\label{sec:disc}

Seeking to expand Bayesian nonparametric methodology for ordinal regression, we have 
presented a fully nonparametric approach to modeling multivariate ordinal responses 
along with covariates. The inferential power of the framework lies in the flexible DP mixture model 
for the latent responses and covariates.
%which allows the data to drive the way in which the covariates affect
%the response. 
The assumption of random covariates is appropriate for many 
problems, and modeling the covariates along with the latent responses accounts for dependence 
or interactions among the covariates. This also allows for inference on 
functionals of the covariate distribution. By establishing its KL support, we have shown that 
the prior probability model can accommodate 
{general}
mixed ordinal-continuous distributions, without 
imputing cut-off points or restricting the covariance matrix of the normal kernel for the DP 
mixture model. From a practical point of view, this is a particularly appealing feature of the 
modeling approach relative to the multivariate probit model and related semiparametric 
extensions.

The multivariate normal mixture kernel can accommodate any type of continuous covariates
(using transformation as needed). Discrete ordinal covariates can also be included by introducing 
latent continuous variables; this was implemented in the example of Section \ref{sec:countries}, 
in which the continuous covariate income was recorded on an ordinal scale. In order to handle 
discrete nominal covariates, the kernel can be modified adding appropriate components to the 
multivariate normal density, using either a marginal or conditional specification \citep[e.g.,][]{taddy}.

The version of the multivariate probit model discussed in the Introduction,
and the setting we consider for our model, involves a common vector of covariates
$\boldsymbol{X}=$ $(X_{1},...,X_{p})$ for each response vector $\boldsymbol{Y}$. 
That is, the covariates are not specific to particular response variables,
but rather $(\boldsymbol{Y},\boldsymbol{X})$ arises as a multivariate vector. 
An alternative version of the probit model involves $p_j$ covariates 
$(X_{j,1},\dots,X_{j,p_j})$ specific to response variable $Y_j$. 
This regression setting was described for multivariate continuous responses 
by \citet{tiao}, and this is the version of the multivariate binary
probit model considered in \citet{chibgreen2}.

Scenarios which make use of response specific covariates fall
broadly into two categories. The first consists of problems in which
only a portion of the covariate vector is thought to affect a
particular response, but there may be some overlap in the subset of
covariates which generate the responses. \citet{chibgreen2} considered
a voting behavior problem of this kind in which the first of two
responses was assumed to be generated by a subset of the covariates
associated with the second response. This data structure can also be 
accommodated by modeling all covariates $\boldsymbol{X}$ jointly with 
$\boldsymbol{Y}$, and conditioning on the relevant subset of 
$\boldsymbol{X}$ in the regression inferences.

The other type of data structure which is occasionally handled with
a multivariate regression model with response specific
covariates involves univariate ordinal responses that are related in
a hierarchical/dynamic fashion. For instance,
%For instance, \citet{tiao} mention an example in
%which each response $Y_j$ corresponds to a particular company $j$, and
%therefore $\boldsymbol{X}_j$ is company-specific. 
\citet{chibgreen2} illustrate their model with the commonly used Six Cities
data, in which $\boldsymbol{Y}=(Y_{1},\dots,Y_{4})$ represents
wheezing status at ages $7$ through $10$. 
Such settings are arguably more naturally approached through
hierarchical/dynamic modeling. Indeed, a possible extension of the
methodology developed here involves dynamic modeling for ordinal regression
relationships, such that at any particular time point a unique regression
relationship is estimated in a flexible fashion, while dependence is
incorporated across time.

\section*{Acknowledgments}

The work of the first author was supported by the National Science Foundation under 
award SES 1131897. The work of the second author was supported in part by the 
National Science Foundation under award DMS 1310438.
The authors wish to thank a reviewer and the editor for constructive feedback 
and for comments that improved the presentation of the material in the paper.

\bigskip
%\begin{center}
%{\large\bf Appendices}
%\end{center}

\appendix
\section*{Appendix A: Proof of Lemma 1}

Under the normal kernel, 
$\text{N}(\boldsymbol{z},\boldsymbol{x} \mid \boldsymbol{\mu},\boldsymbol{\Sigma})$,
of the DP mixture model for the latent responses, $\boldsymbol{z}=$
$(z_{1},...,z_{k})$, and covariates, $\boldsymbol{x}$, the kernel of the implied mixture model 
for the ordinal responses, $\boldsymbol{y}=$ $(y_{1},...,y_{k})$, and covariates is given by
$k(\boldsymbol{y},\boldsymbol{x} \mid \boldsymbol{\mu},\boldsymbol{\Sigma})=$
$\int_{\gamma_{k,y_{k}-1}}^{\gamma_ {k,y_k}}\cdot\cdot\cdot \int_{\gamma_{1,y_{1}-1}}^{\gamma_{1,y_1}}
\text{N}(\boldsymbol{z},\boldsymbol{x} \mid \boldsymbol{\mu},\boldsymbol{\Sigma}) \text{d}z_1 \dots \text{d}z_k$.
We assume fixed cut-off points, and $y_j\in \{1,\dots,C_j\}$ with $C_j>2$, for $j=1,\dots,k$.

We establish likelihood identifiability for parameters $\boldsymbol{\mu}$ and 
$\boldsymbol{\Sigma}$ in 
$k(\boldsymbol{y},\boldsymbol{x} \mid \boldsymbol{\mu},\boldsymbol{\Sigma})$.
That is, from
\begin{equation}\label{eqn:ident}
k(\boldsymbol{y},\boldsymbol{x} \mid \boldsymbol{\mu}_1,\boldsymbol{\Sigma}_1) = 
k(\boldsymbol{y},\boldsymbol{x} \mid \boldsymbol{\mu}_2,\boldsymbol{\Sigma}_2),
\end{equation}
for all $(\boldsymbol{y},\boldsymbol{x})$, with $y_j\in \{1,\dots,C_j\}$ and 
$\boldsymbol{x}\in \mathbb{R}^p$, we will obtain $\boldsymbol{\mu}_1=\boldsymbol{\mu}_2$ 
and $\boldsymbol{\Sigma}_1=\boldsymbol{\Sigma}_2$.

Marginalizing over $\boldsymbol{y}$ both sides of (\ref{eqn:ident}),
we obtain $\text{N}(\boldsymbol{x} \mid \boldsymbol{\mu}_1^x,\boldsymbol{\Sigma}_1^{xx})=$
$\text{N}(\boldsymbol{x} \mid \boldsymbol{\mu}_2^x,\boldsymbol{\Sigma}_2^{xx})$, for all 
$\boldsymbol{x}\in \mathbb{R}^p$, and thus $\boldsymbol{\mu}_1^x=$
$\boldsymbol{\mu}_{2}^{x} \equiv$  $\boldsymbol{\mu}^{x}$, and 
$\boldsymbol{\Sigma}_1^{xx}=$ $\boldsymbol{\Sigma}_2^{xx} \equiv$
$\boldsymbol{\Sigma}^{xx}$. We also have from (\ref{eqn:ident}) that for each $j=1,\dots,k$, 
$k(y_j\mid  \boldsymbol{x},\boldsymbol{\mu}_1,\boldsymbol{\Sigma}_1)=$
$k(y_j\mid  \boldsymbol{x},\boldsymbol{\mu}_2,\boldsymbol{\Sigma}_2)$, for all 
$y_j\in \{1,\dots,C_j\}$ and $\boldsymbol{x}\in \mathbb{R}^p$. Hence,
$\text{Pr}(Y_{j} \leq l \mid \boldsymbol{x},\boldsymbol{\mu}_1,\boldsymbol{\Sigma}_1)=$
$\text{Pr}(Y_{j} \leq l \mid \boldsymbol{x},\boldsymbol{\mu}_2,\boldsymbol{\Sigma}_2)$,
for $l=1,\dots,C_j-1$, and for all $\boldsymbol{x}\in \mathbb{R}^p$. That is, 
\begin{equation}
\label{eqn:univ}
\Phi\left(\frac{\gamma_{j,l}-\mu_1^{z_j}-\boldsymbol{\Sigma}_1^{z_jx}(\boldsymbol{\Sigma}^{xx})^{-1}(\boldsymbol{x} - \boldsymbol{\mu}^{x})}{({\Sigma}_1^{z_jz_j}-\boldsymbol{\Sigma}_1^{z_jx}(\boldsymbol{\Sigma}^{xx})^{-1}\boldsymbol{\Sigma}_1^{xz_j})^{1/2}}\right)=\Phi\left(\frac{\gamma_{j,l}-\mu_2^{z_j}-\boldsymbol{\Sigma}_2^{z_jx}(\boldsymbol{\Sigma}^{xx})^{-1}(\boldsymbol{x}-\boldsymbol{\mu}^{x})}{({\Sigma}_2^{z_jz_j}-\boldsymbol{\Sigma}_2^{z_jx}(\boldsymbol{\Sigma}^{xx})^{-1}\boldsymbol{\Sigma}_2^{xz_j})^{1/2}}\right),
\end{equation}
for all $\boldsymbol{x}\in \mathbb{R}^p$, and for $l=1,\dots,C_j-1$.  
Because $\Phi(\cdot)$ is an increasing function, its arguments in equation (\ref{eqn:univ}) 
must be equal. The resulting equation can be expressed in the form 
$\boldsymbol{a}^{T} \boldsymbol{x} + b = 0$. For this equation to hold true for any 
$\boldsymbol{x}$, we must have $\boldsymbol{a}=\boldsymbol{0}$ and $b=0$, which yield
\begin{equation}
\label{eqn:cond1}\frac{\boldsymbol{\Sigma}_1^{z_jx}}{({\Sigma}_1^{z_jz_j}-\boldsymbol{\Sigma}_1^{z_jx}(\boldsymbol{\Sigma}^{xx})^{-1}\boldsymbol{\Sigma}_1^{xz_j})^{1/2}}=\frac{\boldsymbol{\Sigma}_2^{z_jx}}{({\Sigma}_2^{z_jz_j}-\boldsymbol{\Sigma}_2^{z_jx}(\boldsymbol{\Sigma}^{xx})^{-1}\boldsymbol{\Sigma}_2^{xz_j})^{1/2}},
\end{equation}
and
\begin{equation}
\label{eqn:cond2}
\frac{\gamma_{j,l} - \mu_{1}^{z_j} + \boldsymbol{\Sigma}_1^{z_jx}(\boldsymbol{\Sigma}^{xx})^{-1} \boldsymbol{\mu}^{x}}
{({\Sigma}_1^{z_jz_j}-\boldsymbol{\Sigma}_1^{z_jx}(\boldsymbol{\Sigma}^{xx})^{-1}\boldsymbol{\Sigma}_1^{xz_j})^{1/2}} =
\frac{\gamma_{j,l} - \mu_{2}^{z_j} + \boldsymbol{\Sigma}_{2}^{z_jx}(\boldsymbol{\Sigma}^{xx})^{-1} \boldsymbol{\mu}^{x}}
{({\Sigma}_2^{z_jz_j}-\boldsymbol{\Sigma}_2^{z_jx}(\boldsymbol{\Sigma}^{xx})^{-1}\boldsymbol{\Sigma}_2^{xz_j})^{1/2}},
\end{equation}
for $l=1,\dots,C_j-1$. Using (\ref{eqn:cond1}), (\ref{eqn:cond2}) can be expressed as 
$(\gamma_{j,l}-\mu_1^{z_j})\boldsymbol{\Sigma}_2^{z_jx}=$ 
$(\gamma_{j,l}-\mu_2^{z_j})\boldsymbol{\Sigma}_1^{z_jx}$, for each $j=1,...,k$.
Working with $2$ of these $C_{j} - 1$ equations, the system can be shown to have solution 
$\boldsymbol{\Sigma}_1^{z_jx}=$ $\boldsymbol{\Sigma}_2^{z_jx} \equiv$  $\boldsymbol{\Sigma}^{z_jx}$.
Then, from (\ref{eqn:cond2}) and (\ref{eqn:cond1}), we obtain $\mu_1^{z_j}=$ $\mu_2^{z_j} \equiv$ $\mu^{z_j}$
and ${\Sigma}_1^{z_jz_j}=$ ${\Sigma}_2^{z_jz_j} \equiv$ ${\Sigma}^{z_jz_j}$, respectively. 

Notice that we required $2$ of the $C_{j} - 1$ equations of the form in (\ref{eqn:cond2})
to arrive at this solution. If $C_j=2$ for some $j$, we are unable to identify the full covariance
matrix $\boldsymbol{\Sigma}$. In this case, if we fix ${\Sigma}^{z_jz_j}$, we can identify $\mu^{z_j}$ 
and $\boldsymbol{\Sigma}^{z_jx}$, as in \cite{deyoreo}. Although we do not require free 
cut-offs here due to the flexibility provided by the mixture, if $C_j>3$, the cut-off
points $\gamma_{j,3},\dots,\gamma_{j,C_j-1}$ are also identifiable.

Finally, we must establish identifiability for $\Sigma^{z_jz_{j'}}$, where $j\neq j'$.
Note that (\ref{eqn:ident}) implies $k(y_j,y_{j'} \mid \boldsymbol{\mu}_1,\boldsymbol{\Sigma}_1)=$
$k(y_j,y_{j'} \mid \boldsymbol{\mu}_2,\boldsymbol{\Sigma}_2)$, for any $j,{j'} \in \{1,\dots,k\}$, 
with $j\neq j'$. Hence, for any $y_j\in \{1,\dots,C_j\}$ and 
$y_{j'}\in \{1,\dots,C_{j'}\}$, $\int_{\gamma_{j',y_{j'}-1}}^{\gamma_{j',y_{j'}}} \int_{\gamma_{{j},y_{j}-1}}^{\gamma_{{j},y_{{j}}}} 
\text{N}((z_j,z_{j'})^{T} \mid (\mu^{z_{j}},\mu^{z_{j'}})^{T},\boldsymbol{V}_{1}) \, \text{d}z_j \text{d}z_{j'} $
$=\int_{\gamma_{j',y_{j'}-1}}^{\gamma_{j',y_{j'}}} \int_{\gamma_{{j},y_{j}-1}}^{\gamma_{{j},y_{{j}}}} 
\text{N}((z_j,z_{j'})^{T} \mid (\mu^{z_{j}},\mu^{z_{j'}})^{T},\boldsymbol{V}_{2}) \, \text{d}z_j \text{d}z_{j'}$.
Here, matrices $\boldsymbol{V}_{1}$ and $\boldsymbol{V}_{2}$ have the same diagonal elements 
(given by ${\Sigma}^{z_jz_j}$ and ${\Sigma}^{z_{j'} z_{j'}}$) and off-diagonal element given by 
$\Sigma^{z_j z_{j'}}_{1}$ and $\Sigma^{z_j z_{j'}}_{2}$, respectively. 

%
%\[
%\int_{
%\gamma_{j',y_{j'}-1}}^{\gamma_{j',y_{j'}}}\int_{
%\gamma_{{j},y_{j}-1}}^{\gamma_{{j},y_{{j}}}} \text{N}\left((z_j,z_{j'})^T;(\mu^{z_{j}},\mu^{z_{j'}})^T,\left( \begin{array}{cc}
%\Sigma^{z_jz_j} & \Sigma^{z_iz_{j'}}_1  \\
%\Sigma^{z_iz_{j'}}_1 & \Sigma^{z_{j'}z_{j'}} \end{array} \right)\right) \text{d}z_j \text{d}z_{j'}= 
%\]
%\begin{equation}
%\int_{
%\gamma_{{j'},y_{j'}-1}}^{\gamma_{{j'},y_{j'}}}\int_{
%\gamma_{j,y_j-1}}^{\gamma_{j,y_j}} \text{N}\left((z_j,z_{j'})^T;(\mu^{z_{j}},\mu^{z_{j'}})^T,\left( \begin{array}{cc}
%\Sigma^{z_jz_j} & \Sigma^{z_jz_{j'}}_2  \\
%\Sigma^{z_jz_{j'}}_2 & \Sigma^{z_{j'}z_{j'}} \end{array} \right)\right) \text{d}z_j \text{d}z_{j'}.
%\end{equation}
%
We will therefore obtain $\Sigma^{z_j z_{j'}}_{1} =$ $\Sigma^{z_j z_{j'}}_{2}$, for $j\neq j'$,
if the bivariate normal distribution function is increasing in the correlation 
parameter $\rho$ (assuming, without loss of generality, zero means and unit variances). 
To this end, note that 
$\frac{\partial}{\partial \rho} \text{N}((x,y)^{T} \mid \boldsymbol{0},\boldsymbol{R})=$
$\frac{\partial^{2}}{\partial x \partial y} \text{N}((x,y)^{T} \mid \boldsymbol{0},\boldsymbol{R})$,
where $\boldsymbol{R}$ has unit diagonal elements, and off-diagonal element $\rho$
\citep[e.g.,][]{plackett1954}.
Therefore, 
$\frac{\partial}{\partial \rho} \int_{-\infty}^{a} \int_{-\infty}^{b} 
\text{N}((x,y)^{T} \mid \boldsymbol{0},\boldsymbol{R}) \, \text{d}x \text{d}y =$
$\text{N}( (a,b)^{T} \mid \boldsymbol{0},\boldsymbol{R}) >0$, and thus the result is obtained.
%
%$\int_{-\infty}^{b}\int_{-\infty}^{a}\text{N}((w_1,w_2)^T;(0,0)^T,\boldsymbol{V}) \, \text{d}w_1 \text{d}w_2$ 
%is monotonically increasing in the off-diagonal element $V_{12}$ of covariance matrix $\boldsymbol{V}$,
%for constants $a$ and $b$. This result arises from
%{\small 
%\[\frac{\partial}{\partial V_{12}}\int_{-\infty}^{b}\int_{-\infty}^{a}\text{N}((w_1,w_2)^T;(0,0)^T,
%\boldsymbol{V}) \, \text{d}w_1 \text{d}w_2=\int_{-\infty}^{b}\int_{-\infty}^{a}\frac{\partial}{\partial %V_{12}}\text{N}((w_1,w_2)^T;(0,0)^T,\boldsymbol{V}) \, \text{d}w_1 \text{d}w_2=\]
%\[\int_{-\infty}^{b}\int_{-\infty}^{a}\frac{\partial^2}{\partial w_1\partial w_2}
%\text{N}((w_1,w_2)^T;(0,0)^T,\boldsymbol{V}) \, \text{d}w_1 \text{d}w_2=
%\frac{\partial^2}{\partial w_1\partial w_2}\int_{-\infty}^{b}\int_{-\infty}^{a}
%\text{N}((w_1,w_2)^T;(0,0)^T,\boldsymbol{V}) \, \text{d}w_1 \text{d}w_2\]
%\[=\mathrm{N}((a,b)^T;(0,0)^T,\boldsymbol{V})>0.\]
%}
%

\section*{Appendix B: Kullback-Leibler support for the model}

{We first prove Lemma 2 assuming (as in the statement
of the lemma) that the true ordinal-continuous distribution 
$p_0(\boldsymbol{x},\boldsymbol{y}) \in \mathcal{D^*}$ satisfies
appropriate regularity conditions which imply that the corresponding 
density $f_0(\boldsymbol{x},\boldsymbol{z}) \in \mathcal{D}$ is in the
KL support of the prior. This result is general and it applies to any
prior model for the latent continuous distribution which provides full KL
support to $f_{0}$. 
Next, we provide a specific example for an $f_{0}$
that enables the connection with $p_{0}$ given in (\ref{eqn:mapping}).
And, using Theorem 2 of \cite{wu}, we derive conditions for $p_{0}$
under which the corresponding $f_{0}$ is in the KL support of the DP
location normal mixture prior model, thus validating Lemma 2 for our
model.}

\subsection*{B1. Proof of Lemma 2}

Let $\mathrm{KL}(f_0,f)=$ $\int f_{0}(w) \log(f_{0}(w)/f(w)) \text{d}w$ be
the KL divergence between densities $f_0$ and $f$. Based on the chain rule for relative entropy, 
$$
\mathrm{KL}(f_0(\boldsymbol{x},\boldsymbol{z}),f(\boldsymbol{x},\boldsymbol{z}))
= \mathrm{KL}(f_0(\boldsymbol{x}),f(\boldsymbol{x})) + \mathrm{KL}(f_0(\boldsymbol{z} \mid
\boldsymbol{x}),f(\boldsymbol{z}\mid \boldsymbol{x}))
$$
where the KL divergence between conditional densities is defined as
$\mathrm{KL}(f_0(\boldsymbol{z}\mid \boldsymbol{x}),f(\boldsymbol{z}\mid \boldsymbol{x}))=$
$\int f_{0}(\boldsymbol{x}) \{ \int f_0(\boldsymbol{z}\mid \boldsymbol{x}) 
\log( f_0(\boldsymbol{z}\mid \boldsymbol{x}) / f(\boldsymbol{z}\mid \boldsymbol{x}) ) 
\text{d} \boldsymbol{z} \} \text{d} \boldsymbol{x}$. Hence,
$\mathrm{KL}(f_0(\boldsymbol{x}),f(\boldsymbol{x})) \leq$
$\mathrm{KL}(f_0(\boldsymbol{x},\boldsymbol{z}),f(\boldsymbol{x},\boldsymbol{z}))$,
and thus, for any $\epsilon > 0$, 
$K_\epsilon(f_0(\boldsymbol{x},\boldsymbol{z})) \subseteq$ $K_\epsilon(f_0(\boldsymbol{x})) =$
$\{f(\boldsymbol{x},\boldsymbol{z}):\mathrm{KL}(f_0(\boldsymbol{x}),f(\boldsymbol{x}))<\epsilon\}$.
Using the KL property of the prior model for $f(\boldsymbol{x},\boldsymbol{z})$, 
$\mathcal{P}(K_\epsilon(f_0(\boldsymbol{x}))) \geq \mathcal{P}(K_\epsilon(f_0(\boldsymbol{x},\boldsymbol{z}))) >0$,
such that the prior assigns positive probability to all KL neighborhoods of 
the true covariate density $f_0(\boldsymbol{x})$.

The proof relies on the following inequality for two densities 
$g_1(\boldsymbol{t})$ and $g_2(\boldsymbol{t})$, where $\boldsymbol{t} \in \mathbb{R}^{s}$,
and for general subsets $A_{1},...,A_{s}$ of $\mathbb{R}$,
\begin{equation}
\label{eqn:ineq}
\int_{A_s} \cdot\cdot\cdot \int_{A_1} g_1(\boldsymbol{t})
\log\left(\frac{g_1(\boldsymbol{t})}{g_2(\boldsymbol{t})} \right) \text{d}\boldsymbol{t} \geq
\int_{A_s} \cdot\cdot\cdot \int_{A_1}  g_1(\boldsymbol{t}) \text{d}\boldsymbol{t} \times \log\left(\frac{\int_{A_s}\cdot\cdot\cdot\int_{A_1}g_1(\boldsymbol{t}) \text{d}\boldsymbol{t}}{\int_{A_s}\cdot\cdot\cdot\int_{A_1}g_2(\boldsymbol{t})
\text{d}\boldsymbol{t}}\right).
\end{equation}
To prove the inequality, let $B_{r}=$ $\int_{A_s} \cdot\cdot\cdot \int_{A_1} g_{r}(\boldsymbol{t}) 
\text{d}\boldsymbol{t}$, for $r=1,2$, such that $h_{r}(\boldsymbol{t})=$ $g_{r}(\boldsymbol{t})/B_{r}$,
$r=1,2$, are densities on $A_{1} \times ... \times A_{s}$. Then, the left-hand-side of (\ref{eqn:ineq}) 
can be written as \linebreak
$B_{1} \int_{A_s} \cdot\cdot\cdot \int_{A_1} h_{1}(\boldsymbol{t})
\log\left(\frac{B_{1} h_{1}(\boldsymbol{t})}{B_{2} h_{2}(\boldsymbol{t})} \right) \text{d}\boldsymbol{t}=$
$B_{1} \log(\frac{B_{1}}{B_{2}})$ + 
$B_{1} \int_{A_s} \cdot\cdot\cdot \int_{A_1} h_{1}(\boldsymbol{t})
\log\left(\frac{h_{1}(\boldsymbol{t})}{h_{2}(\boldsymbol{t})} \right) \text{d}\boldsymbol{t} \geq$
$B_{1} \log(\frac{B_{1}}{B_{2}})$, since 
$\int_{A_s} \cdot\cdot\cdot \int_{A_1} h_{1}(\boldsymbol{t})
\log\left( h_{1}(\boldsymbol{t})/h_{2}(\boldsymbol{t}) \right) \text{d}\boldsymbol{t}$
is the KL divergence for densities $h_{1}$ and $h_{2}$.

Now, consider density $f(\boldsymbol{x},\boldsymbol{z}) \in K_{\epsilon/2}(f_0(\boldsymbol{x},\boldsymbol{z}))$. 
By the chain rule, $\mathrm{KL}(f_0(\boldsymbol{x}),f(\boldsymbol{x}))<\epsilon/2$, and 
$\mathrm{KL}(f_0(\boldsymbol{z}\mid \boldsymbol{x}),f(\boldsymbol{z}\mid \boldsymbol{x}))<\epsilon/2$. 
We apply (\ref{eqn:ineq}) with $s=k$, $g_1(\boldsymbol{t})=$ $f_0(\boldsymbol{z}\mid \boldsymbol{x})$, 
$g_2(\boldsymbol{t})=$ $f(\boldsymbol{z}\mid \boldsymbol{x})$, and $A_j=$ 
$(\gamma_{j,y_j-1},\gamma_{j,y_j})$, for $j=1,\dots,k$. This yields
\begin{equation}
\label{ineq-2}
\int_{\gamma_{k,y_k-1}}^{\gamma_{k,y_k}} \cdot\cdot\cdot \int_{\gamma_{1,y_1-1}}^{\gamma_{1,y_1}}
f_{0}(\boldsymbol{z}\mid  \boldsymbol{x})  
\log\left(\frac{f_0(\boldsymbol{z}\mid \boldsymbol{x})}{f(\boldsymbol{z}\mid \boldsymbol{x})}\right) 
\text{d}\boldsymbol{z}  \geq  p_0(\boldsymbol{y} \mid \boldsymbol{x})
\log\left( \frac{p_0(\boldsymbol{y} \mid  \boldsymbol{x})}{p^*(\boldsymbol{y}\mid \boldsymbol{x})}
\right)
\end{equation}
for any configuration of values $\boldsymbol{y}=$ $(y_{1},...,y_{k})$ for the ordinal responses.
Here, $p^*(\boldsymbol{y} \mid \boldsymbol{x})=$
$\int_{\gamma_{k,y_k-1}}^{\gamma_{k,y_k}}\cdot\cdot\cdot\int_{\gamma_{1,y_1-1}}^{\gamma_{1,y_1}}
f(\boldsymbol{z} \mid \boldsymbol{x}) \text{d}\boldsymbol{z}$, such that
$p^*(\boldsymbol{x},\boldsymbol{y})=$ $f(\boldsymbol{x}) p^*(\boldsymbol{y} \mid \boldsymbol{x})$.
Next, we sum both sides of (\ref{ineq-2}) over $\boldsymbol{y}$, then multiply both sides
by $f_0(\boldsymbol{x})$, and finally integrate both sides over $\boldsymbol{x}$, to obtain
\[
\int_{\mathbb{R}^{p}} f_0(\boldsymbol{x}) \int_{\mathbb{R}^{k}} f_0(\boldsymbol{z} \mid \boldsymbol{x})
\log\left(\frac{f_0(\boldsymbol{z}\mid \boldsymbol{x})}{f(\boldsymbol{z}\mid \boldsymbol{x})}\right) 
\text{d}\boldsymbol{z} \text{d}\boldsymbol{x} \geq 
\int_{\mathbb{R}^{p}} f_0(\boldsymbol{x})\sum_{y_k=1}^{C_k}\cdot\cdot\cdot \sum_{y_1=1}^{C_1} 
p_0(\boldsymbol{y}\mid \boldsymbol{x})  \log\left(\frac{p_0(\boldsymbol{y}\mid \boldsymbol{x})}
{p^*(\boldsymbol{y}\mid \boldsymbol{x})}\right) \text{d}\boldsymbol{x}
\]
that is, $\mathrm{KL}(f_0(\boldsymbol{z}\mid \boldsymbol{x}),f(\boldsymbol{z}\mid \boldsymbol{x})) \geq$
$\mathrm{KL}(p_0(\boldsymbol{y}\mid \boldsymbol{x}),p^*(\boldsymbol{y}\mid \boldsymbol{x}))$.
Since $\mathrm{KL}(f_0(\boldsymbol{z}\mid \boldsymbol{x}),f(\boldsymbol{z}\mid \boldsymbol{x}))<\epsilon/2$,
we have $\mathrm{KL}(p_0(\boldsymbol{y}\mid \boldsymbol{x}),p^*(\boldsymbol{y}\mid \boldsymbol{x})) < \epsilon/2$,
which in conjunction with $\mathrm{KL}(f_0(\boldsymbol{x}),f(\boldsymbol{x}))<\epsilon/2$, further implies
$\mathrm{KL}(p_0(\boldsymbol{x},\boldsymbol{y}),p^*(\boldsymbol{x},\boldsymbol{y}))<\epsilon$, by the chain rule. 
We have thus obtained
\[
K_{\epsilon/2}(f_0(\boldsymbol{x},\boldsymbol{z})) \subseteq K_{\epsilon/2}(p_0(\boldsymbol{y}\mid \boldsymbol{x}))
\,\,\,\,\,\,\,   \text{and}   \,\,\,\,\,\,\,
K_{\epsilon/2}(f_0(\boldsymbol{x},\boldsymbol{z})) \subseteq K_{\epsilon}(p_0(\boldsymbol{x},\boldsymbol{y}))
\]
where $K_\epsilon(p_0(\boldsymbol{x},\boldsymbol{y}))=$ $\{f(\boldsymbol{x},\boldsymbol{z}):
\mathrm{KL}(p_0(\boldsymbol{x},\boldsymbol{y}),p^*(\boldsymbol{x},\boldsymbol{y}))<\epsilon\}$,
from which the result emerges using the KL property of the prior model for $f(\boldsymbol{x},\boldsymbol{z})$.

\subsection*{B2. Density $f_{0}$ and conditions for the true $p_{0}$}

We first show that there exists at least one $f_0(\boldsymbol{x},\boldsymbol{z}) \in \mathcal{D}$ 
for any $p_0(\boldsymbol{x},\boldsymbol{y}) \in \mathcal{D^*}$, as defined in (\ref{eqn:mapping}). 
Consider $y_j \in \{1,\dots,C_j\}$, for $j=1,...,k$, and define 
\begin{equation}
\label{true_p0_f0}
f_0(\boldsymbol{x},\boldsymbol{z}) = \sum_{y_{1}} \cdot\cdot\cdot \sum_{y_{k}}
\frac{p_0(\boldsymbol{x},y_1,\dots,y_k) 
\prod_{j=1}^k 1_{(\gamma^*_{j,y_{j}-1},\gamma^*_{j,y_j}]}(z_j)}{\prod_{j=1}^k(\gamma^*_{j,y_j}-\gamma^*_{j,y_j-1})}
\end{equation}
where $\gamma^*_{j,l}=\gamma_{j,l}$ if $l\in \{1,\dots,C_j-1\}$, $\gamma^*_{j,0}=b_j$, 
and $\gamma^*_{j,C_j}=d_j$, with $-\infty<b_j<\gamma_{j,1}$ and $\gamma_{j,C_{j}-1}<d_j<\infty$, 
for $j=1,\dots,k$. Then, $f_0(\boldsymbol{x},\boldsymbol{z})$ satisfies the relationship 
in (\ref{eqn:mapping}).

{The final item needed to show that Lemma 2 is
applicable to our model is to develop conditions for the true $p_{0}$
under which the corresponding $f_{0}$ in (\ref{true_p0_f0}) satisfies
(\ref{eqn:KLprop}). To this end, we use Theorem 2 of \cite{wu}, which
provides nine conditions (B1 to B9) on the true density $f_{0}$ and on the
kernel of a nonparametric mixture model such that the KL property is 
satisfied. Conditions B1, B2, B3 and B9 involve the mixture
kernel, and as discussed in \cite{wu}, are satisfied by the product
normal kernel. Condition B8 involves the prior for the 
mixing distribution and is valid for the DP prior.}

{
Conditions A, B, C and D stated below for the true $p_{0}$ are
sufficient for the remaining conditions in Theorem 2 of \cite{wu} to hold.
For simpler notation, we consider the special case of
the model with a single ordinal response and a single covariate, such
that $p_{0}(x,y)$ is a distribution on $\mathbb{R} \times \{ 1,...,C \}$,
with $C \geq 2$. The conditions can be extended to the general setting 
with multiple ordinal responses and covariates.}

\vspace{0.15cm}

\noindent
{
{\bf A.} There exists $0 < M < \infty$ such that 
$0 < p_{0}(x,y) \leq M$, for all $(x,y) \in \mathbb{R} \times \{ 1,...,C \}$.
\\
{\bf B.} 
$| \int p_{0}(x,y) \log( p_{0}(x,y)/W ) \, \text{d}x | < \infty$, for all 
$y \in \{ 1,...,C \}$, and for any $W>0$.
\\
{\bf C.} There exists $\delta > 0$ such that 
$\int p_{0}(x,y) \log( p_{0}(x,y)/\varphi_{\delta}(x,y) ) \, \text{d}x
< \infty$, for all $y \in \{ 1,...,C \}$, where $\varphi_{\delta}(x,y)=$
$\inf_{ |x'-x| < \delta } p_{0}(x',y)$.
\\
{\bf D.} There exist $\eta > 0$ and $d > 0$ such that
$\int p_{0}(x,y) (x^{2} + d^{2})^{\eta} \, \text{d}x < \infty$ and 
$\int p_{0}(x,y) x^{2} (x^{2} + d^{2})^{\eta} \, \text{d}x < \infty$, 
for all $y \in \{ 1,...,C \}$. In addition, 
$\int p_{0}(x,y) \{ (x-a)^{2}/(2 v^{2}) \} \, \text{d}x < \infty$, for all
$y \in \{ 1,...,C \}$, and for any $a \in \mathbb{R}$ and $v>0$.
}

\vspace{0.1cm}

\noindent
{It is straightforward to verify that B4, B5, B6 and B7 in Theorem 2 of
\cite{wu} are satisfied by $f_{0}$ in (\ref{true_p0_f0}) under
conditions A, B, C and D, respectively. The only additional assumption 
needed is that the cut-off points take positive values, which is
without loss of generality under our model that uses fixed cut-offs.}

%
%To ensure the KL property is satisfied for the proposed model, we also include
%in Appendix B2 a set of conditions for $p_{0}$ under which the specific
%corresponding $f_{0}$ satisfies (\ref{eqn:KLprop}), in particular, it satisfies 
%the general conditions for the KL property given in Theorem 2 of \cite{wu}.
%
%Next, we provide a specific example for an $f_{0}$
%that enables the connection with $p_{0}$ given in (\ref{eqn:mapping}).
%And, using Theorem 2 of \cite{wu}, we derive conditions for $p_{0}$
%under which the corresponding $f_{0}$ is in the KL support of the DP
%location normal mixture prior model
%

\section*{Supplementary Materials}
All supplemental files listed below are contained in a single .zip
file (supplementary-files.zip) and can be obtained via a single download.

\begin{description}
\item[Ozone data:] ozone data file. (ozone.csv)
\item[Credit rating data:] credit rating data. (credit.dat)
\item[Countries data:] S\&P ratings of countries. (countries.txt)
\item[Multirater data:] data used in the ratings of student essays example. (essay.csv)
\item[Code for ozone data:] Code for implementing ordinal regression model on ozone data. (ozone-code-paper, R file)
\item[Code for credit rating data:] Code for implementing ordinal regression model on credit rating data. (credit-code-paper, R file)

\item[Code for multirater data:] Code for implementing ordinal regression model on multirater data. (multirater-code-paper, R file)
\item[MCMC supplement:] Additional material related to MCMC and computation. (supplementary-mcmc.pdf) 
\end{description}

\singlespacing
%\nocite{*}
\bibliographystyle{asa}
\bibliography{ordregpaperbib}

\end{document}